\documentclass[twocolumn]{aastex631}

\newcommand{\omcen}{$\omega$~Cen}

\usepackage{tabularx}
\usepackage{amsmath}
\usepackage{amstext}
\usepackage{apjfonts}
\usepackage{graphicx}

\usepackage{amssymb}
\usepackage{latexsym}
\usepackage{longtable}

\shorttitle{Wide View of NGC~2808}
\shortauthors{Johnson et al.}
\graphicspath{{./}{figures/}}

\begin{document}

\title{A Wide View of the Galactic Globular Cluster NGC~2808: Red Giant and Horizontal Branch Star
Spatial Distributions}

\author[0000-0002-8878-3315]{Christian I. Johnson}
\affiliation{Space Telescope Science Institute \\
3700 San Martin Drive \\
Baltimore, MD 21218, USA}

\author[0000-0002-0882-7702]{Annalisa Calamida}
\affiliation{Space Telescope Science Institute \\
3700 San Martin Drive \\
Baltimore, MD 21218, USA}

\author[0000-0002-6650-3757]{Justin A. Kader}
\affiliation{Department of Physics and Astronomy, 4129 Frederick Reines Hall \\
University of California, Irvine \\
Irvine, CA  92697, USA}

\author{Ivan Ferraro}
\affiliation{INAF - Osservatorio Astronomico di Roma \\
Via Frascati 33 \\
00040, Monte Porzio Catone, Italy}

\author[0000-0002-3007-206X]{Catherine A. Pilachowski}
\affiliation{Department of Astronomy, Indiana University \\
SW 319, 727 East 3rd Street \\
Bloomington, IN  47405, USA}

\author[0000-0002-4896-8841]{Giuseppe Bono}
\affiliation{Department of Physics, Università di Roma Tor Vergata \\
Via della Ricerca Scientifica 1 \\
00133, Roma, Italy}

\author[0000-0002-2386-9142]{Alessandra Mastrobuono-Battisti}
\affiliation{GEPI, Observatoire de Paris, PSL Research University, CNRS \\
Place Jules Janssen \\
F-92190 Meudon, France}

\author[0000-0002-4410-5387]{Armin Rest}
\affiliation{Space Telescope Science Institute \\
3700 San Martin Drive \\
Baltimore, MD 21218, USA}

\author[0000-0001-6455-9135]{Alfredo Zenteno}
\affiliation{Cerro Tololo Inter-American Observatory \\
NSF's National Optical-Infrared Astronomy Research Laboratory \\ 
Casilla 603 La Serena, Chile}

\author[0000-0001-8075-3149]{Alice Zocchi}
\affiliation{University of Vienna 
Universit\"atsring 1 \\
1010 Vienna, Austria}



\begin{abstract}
Wide-field and deep \emph{DECam} multi-band photometry, combined with \emph{HST} data for the core of the Galactic globular cluster NGC~2808, allowed us to study the distribution of various stellar sub-populations and stars in different evolutionary phases out to the cluster tidal radius.
We used the $C_{ugi} = (u-g) - (g-i)$ index to identify three chemically distinct sub-populations along the red giant branch and compared their spatial distributions.  The most light-element enriched sub-population (P3) is more centrally concentrated; however, it shows a more extended distribution in the external regions of the cluster compared to the primordial (P1) and intermediate (P2) composition populations. Furthermore, the P3 sub-population centroid is off-center relative to those of the P1 and P2 groups. 
We also analyzed the spatial distribution of horizontal branch stars and found that the relative fraction of red horizontal branch stars increases for radial distances larger than $\approx$ 1.5\arcmin while that of the blue and hotter stars decreases. These new observations, combined with literature spectroscopic measurements, suggest that the red horizontal branch stars are the progeny of all the stellar sub-populations in NGC~2808, i.e. primordial and light-element enhanced, while the blue stars are possibly the result of a combination of the "hot-flasher" and the "helium-enhanced" scenarios.
A similar distribution of different red giant branch sub-populations and horizontal branch stars was also found for the most massive Galactic globular cluster, \omcen, based on combined \emph{DECam} and \emph{HST} data, which suggests the two may share a similar origin.

\end{abstract}



\section{Introduction} \label{sec:intro}

NGC~2808 is one of the most massive ($M$ = 8.5$\times 10^5$ $M_{\odot}$, \citealt{mclaughlin2005}) Galactic globular clusters (GGC) 
and a very peculiar object. \textit{Hubble Space Telescope} (\emph{HST}) photometric investigations revealed that the cluster main-sequence (MS) splits into a blue, an intermediate, and a red sequence. Furthermore, NGC~2808 shows an extended horizontal branch (HB) with distinct components: a red HB (RHB) and a blue tail divided into three groups \citep{sosin1997, Bedin2000}.
Therefore, it was suggested that NGC~2808 experienced multiple episodes of 
star formation, each with varying levels of helium enrichment and with the bluest MS representing the most enhanced population \citep{Piotto07, Milone12, Milone15}. 
The helium enrichment could also explain the observed HB morphology \citep{dantona2004,Dantona05, Lee05}. Table~\ref{table:1} summarizes the basic properties of NGC~2808.

High-resolution spectroscopic measurements of $\sim$ 140 red giant branch (RGB) stars in the cluster indicate that, while expected in the case of different star formation episodes, no spread in iron content is present in NGC~2808 ($[Fe/H]$ = $-$1.192$\pm$0.004$\pm$0.034, \citealt{Carretta15}). This already puzzling picture is complicated by the fact that the RGB evolutionary phase splits into five stellar populations with different light-element abundances \citep{Carretta15}.

\citet{iannicola2009} combined \emph{HST} and ground-based data covering a field-of-view (FoV) of $\sim$ 15\arcmin$\times$15\arcmin~ centered on NGC~2808 to show that the relative fraction of cool (red) and hot (blue tail) HB stars is constant from the center to the outskirts of the cluster. This result supports the lack of radial differentiation among NGC~2808 stellar populations with possible distinct helium abundances. On the other hand, \citet{simioni2016} showed that the intermediate and blue MS stars in NGC~2808, supposedly more helium-enhanced, are more centrally concentrated compared to red MS stars, at least out to radial distances of 8$\arcmin$ 
($\sim$1/3 of the tidal radius).

A proper motion study based on \emph{HST} data from \citet{bellini20152015} showed that the three MSs in NGC~2808 also have different kinematic behavior. At the outermost distance probed, r $\approx$ 1.5\arcmin, the velocity distribution of the intermediate and the blue MS stars is radially anisotropic, but it is isotropic for the stars belonging to the red MS. These findings might indicate the diffusion towards the cluster outskirts of the supposedly helium-enhanced populations, initially more concentrated. 
According to model predictions of the formation of GGCs with multiple stellar populations, the second generation of stars should form in the inner regions \citep{Dercole08, Bastian13}: GGCs with a long relaxation time, such as NGC~2808 and \omcen, should still show remnants of this initial spatial segregation. 

A consensus on the origin of the different stellar sub-populations in NGC~2808 has not yet been reached. All previous investigations are based either on data for a few small fields across the center of the cluster (\emph{HST}) or for a FoV covering about half the tidal radius (ground). There is now the need for a deep photometric study covering the entire cluster (tidal radius r$_t \sim$ 22\arcmin, see Table~\ref{table:1}, \citealt{deboer2019}), with the precision necessary to enable the identification of the different RGB and HB groups. 

In order to achieve this goal, we combined \emph{HST} data for the core with deep $ugri$ \emph{DECam} photometry for NGC~2808. \emph{DECam} is a wide-field imager covering a 3 square degree sky FoV (Fig.~\ref{fig:density_plot}).  The high photometric quality and wide field covered by the combined \emph{DECam + HST} photometric catalogs enables us to identify different RGB groups and to study their spatial distribution from the center to the cluster tidal radius. We also analyzed the distribution of red and blue HB stars and compared it to that of RGB, AGB, and MS stars. 

Therefore, \emph{DECam} photometry of NGC~2808 allowed us for the first time to investigate the spatial distribution of the different RGB sub-populations and different evolutionary phases from the core to the tidal radius, and to discover their peculiarities.


\begin{deluxetable}{lrc}
\tablecaption{Positional, photometric and structural parameters of the 
Galactic Globular Cluster NGC~2808\label{table:1}}
\tablehead{
\colhead{Parameter}&
\colhead{         }&   
\colhead{Ref.\tablenotemark{a}} 
}
\startdata
$\alpha$ (J2000)                          &  138.0071          &     1    \\  
$\delta$ (J2000)                          &  -64.8645          &     1    \\  
$\mu_{\alpha}$ (2015)                     &  0.994$\pm$0.024 &     2    \\  
$\mu_{delta}$ (2015)                      &  0.273$\pm$0.024 &     2    \\  
$M_V$ (mag)\tablenotemark{a}              &  -9.4              &     3    \\  
$r_c$ (arcmin)\tablenotemark{b}           &  0.26              &     4	  \\  
$r_h$ (arcmin)\tablenotemark{c}           &  0.86              &     5	  \\  
$r_t$ (arcmin)\tablenotemark{d}           &  21.97             &     5	  \\  
$log(t_h)$ \tablenotemark{e}              & 8.9                &     6    \\
$E(B-V)$\tablenotemark{f}                 & 0.19$\pm$0.03      &     7    \\  
$\mu_0$ (mag)\tablenotemark{g}            & 15.05              &     3    \\  
\enddata 
\tablenotetext{a}{References: 1) \citet{Gaia18_GCs}; 2) \citet{vasiliev21}; 3) \citet{harris2010};
4) \citet{trager1995}; 5) \citet{deboer2019} ; 6) \citet{mclaughlin2005} ; 
7) \citet{schlafly2011}
$^a$ Total Visual magnitude.  
$^b$ Core radius. 
$^c$ Half-mass radius.  
$^d$ Tidal radius.  
$^e$ Log of relaxation time.
$^f$ Reddening.  
$^g$ True distance modulus.
}   
\end{deluxetable}

\section{Observations and Data Analysis} \label{sec:observations}
We collected a set of 207 $ugri$ \emph{DECam}   images for NGC~2808 with NOAO proposals 2014A-0327, 2014B-0378, 2015B-0307, 2016A-0189, 2016A-0191, 2016B-0301, and 2017B-0279 (PI: Rest).
In particular, we observed a couple of dozen deep ($t_{exp}$ = 600s) $u$-band images in good observing conditions in January 2018, reaching a full-width half maximum of $\approx$ 1-1.2\arcsec~ on the images. The $u$-band photometry, which is sensitive to both effective temperature and metallicity, is critical for the color-color-magnitude method used here to separate cluster and field stars \citep{Calamida17,calamida2020}.  Moreover, the $u$ filter is also fundamental to define the $C_{ugi}$ index \citep{Monelli13} that we use to separate RGB stars with different light element abundances.

A set of standard fields from SDSS Stripe82 were also observed each night utilizing 15s exposures for the $gri$-bands and 30s exposures for the $u$-band.  A total of 36 Stripe 82 images were obtained at varying air masses for each filter.  These data were used to transform the instrumental magnitudes onto the \emph{DECam}   natural system (see Section \ref{subsec:pipeline}).

A small number of additional, deep \emph{DECam}   exposures of NGC~2808 were also downloaded from the NOIRLab Astro Data Archive\footnote{The NOIRLab Astro Data Archive can be accessed at: https://astroarchive.noirlab.edu/.} to supplement the data described above.  These included 5 $gr$-band images from program 2013B-0615 (PI: Carballo-Bello) ranging between 200-300s of integration, 4 $ugri$-band images from program 2013A-9999 (PI: Walker), each with 200s of integration, and 12 $ugi$-band images from program 2012B-0001 (PI: Frieman) spanning 200-500s of integration.  Typical seeing was $\sim$ 1$\arcsec$ for the 2013A-9999 data, 1.5$\arcsec$ for the 2013B-0615 data, and 1.1$\arcsec$ for the 2012B-0001 data.  

We did not download any accompanying standard fields for the archival data, and instead transferred the photometric calibration from our fields onto the archival data using stars in common between both sets.  A log of all observations is presented in Table \ref{table:2}.

\begin{table*}
\footnotesize
\caption{Log of NGC~2808 \emph{DECam}   Observations}
\begin{tabular}{cccccccccc}
\hline
File    &   Obs. Date   &   Proposal    &   PI  &   Filter  &   Exp. Time   &   RA  &   DEC &   Seeing  &   Airmass \\
    &   &   &   &   &   (sec)   &   (degrees)   &   (degrees)   &   (arc sec.)  &   \\
\hline
c4d\_160303\_050811\_ooi\_i\_v1 & 2016-03-03 & 2016A-0189 & Rest & i & 15 & 8:43:00.77 & 0:00:02.3 & 1.122 & 1.430 \\
c4d\_160303\_050856\_ooi\_r\_v1 & 2016-03-03 & 2016A-0189 & Rest & r & 15 & 8:43:00.50 & 0:00:05.4 & 1.281 & 1.430 \\
c4d\_160303\_050941\_ooi\_g\_v1 & 2016-03-03 & 2016A-0189 & Rest & g & 15 & 8:43:00.69 & 0:00:01.8 & 1.459 & 1.440 \\
c4d\_160303\_051040\_ooi\_u\_v1 & 2016-03-03 & 2016A-0189 & Rest & u & 30 & 8:43:00.83 & 0:00:06.4 & 1.604 & 1.440 \\
c4d\_160303\_052102\_ooi\_r\_v1 & 2016-03-03 & 2016A-0189 & Rest & r & 15 & 14:42:00.34 & -0:04:27.7 & 1.267 & 1.800 \\
c4d\_160303\_052144\_ooi\_i\_v1 & 2016-03-03 & 2016A-0189 & Rest & i & 15 & 14:42:00.03 & -0:04:28.3 & 1.105 & 1.800 \\
c4d\_160303\_052227\_ooi\_g\_v1 & 2016-03-03 & 2016A-0189 & Rest & g & 15 & 14:42:00.22 & -0:04:24.2 & 1.404 & 1.790 \\
c4d\_160303\_052326\_ooi\_u\_v1 & 2016-03-03 & 2016A-0189 & Rest & u & 30 & 14:42:00.14 & -0:04:24.9 & 1.469 & 1.780 \\
\hline
\multicolumn{10}{c}{The full version of this table is provided in electronic form.}
\end{tabular}
\label{table:2}
\end{table*}

We also downloaded an \emph{HST} catalog for NGC~2808 from the "Hubble Space telescope UV legacy survey of galactic globular clusters" \citep[GO13297, PI: Piotto]{Piotto15, nardiello2018}. This catalog includes photometry collected with the Wide Field Camera 3 (WFC3) blue filters $F275W, F336W,$ and $F438W$ combined with ACS photometry in $F606W$ and $F814W$ by \citet{sarajedini2007}, and covers the central 1-2$\arcmin$ of NGC~2808\footnote{The \emph{HST} data can be obtained from the Mikulski 
Archive for Space Telescopes (MAST) at the Space Telescope Science Institute by using: 
\dataset[DOI: 10.17909/18ex-q697]{http://dx.doi.org/10.17909/18ex-q697}.}.

\begin{figure*}[t!]
\includegraphics[width=\textwidth]{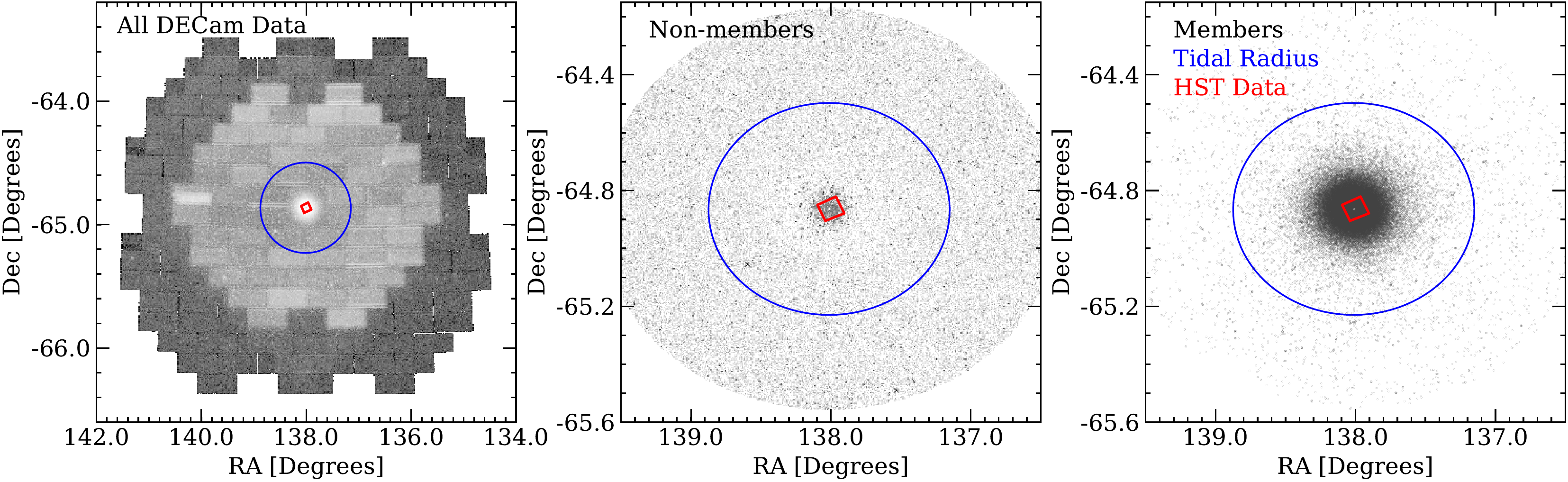}
\caption{The left panel shows a source density map for all \emph{DECam} data utilized in this project, with NGC~2808 at the center of the field.  The middle and right panels show similar source density maps for the inner 40$\arcmin$ region around NGC~2808 that was used to separate non-member (middle) and member (right) stars.  The blue circle in each panel highlights the tidal radius of 21.97$\arcmin$ adopted from \citet{deboer2019}, and the red box illustrates the location of archival \emph{HST}
observations that were used to fill a coverage gap in the very crowded cluster core (r $<$ 1.5$\arcmin$).  Note that even though some stars observed near the cluster
core with \emph{DECam} were identified as non-members, most of these objects are inside the \emph{HST} footprint for which we assumed 100$\%$ membership rates.}
\label{fig:density_plot}
\end{figure*}

The left panel of Fig.~\ref{fig:density_plot} shows a density map of all \emph{DECam}   fields used in this project, and indicates that the exposures extend well beyond the cluster tidal radius.  However, the middle and right panels show that the \emph{DECam}   observations have a hole near the cluster core due to the extreme crowding.  Fortunately, this region is completely covered by the \emph{HST} observations, which include both the ultra-violet (UV) bands necessary for identifying stars with different light element abundances and the very high spatial resolution required to separate and photometer stars in the cluster core.

\subsection{Photometry Pipeline, Calibration, and Catalog Preparation} \label{subsec:pipeline}
The pipeline for generating intermediate catalogs with photometry, astrometry, errors, and associated quality flags closely followed the methods outlined in \citet{Johnson2020}.  To briefly summarize, the full focal plane \emph{DECam}   images, which were pre-processed with the \emph{DECam}   Community Pipeline \citep{Valdes14}, were separated into individual CCD files, and then further partitioned based on filter, data set origin, and science/calibration status.  All exposures were processed independently with DAOPHOTIV/ALLSTAR \citep{Stetson87} on the science servers at the Space Telescope Science Institute, using an "embarrassingly parallel" code implementation.  Both the science and calibration exposures were processed using quadratically varying point spread functions (PSF) along with a minimum of 25 and 10 PSF reference stars, respectively.  Three "fit and subtract" loops were run on each exposure to identify additional faint stars and those in highly crowded fields.

A final processing step subtracted all photometered objects from each exposure except the bright and relatively isolated stars used for PSF fitting.  DAOPHOT's aperture photometry routine was run on these objects, in the subtracted images only, using a set of 12 sky apertures ranging from about 3 to 40 pixels in radius.  The aperture photometry tables were then used to calculate growth-curves via the DAOGROW algorithm from \citet{Stetson90}, and the resulting aperture corrections were applied to the PSF photometry values in all images.

The $r$-band was selected as the astrometric filter since it was observed every night and is in the middle of the wavelength range spanned by our observations.  Following the methods outlined in \citet[][see their Section 3.3.1]{Johnson2020}, we combined all of the $r$-band positions into a single table and generated a catalog of unique sources.  Using a search threshold of 1$\arcsec$, all objects detected in other bands were mapped onto the unique source table.  The final list of unique sources totaled approximately 2$\times$10$^{6}$ objects.  A database was then created that linked all exposure and metadata (e.g., observation date, airmass, etc.) for each unique object so that the photometry could be merged.

Before merging the photometry, one exposure in each band was selected to serve as the zero-point reference frame.  We only selected references frames taken on the same nights as our SDSS Stripe 82 calibration fields.  Median offset values were calculated between the reference frames and all other exposures of the same band using overlapping stars with magnitudes ranging from 14-18, after applying the necessary airmass corrections.  Mean ($\sigma$) image-to-image zero-point offsets for the $ugriz$ bands relative to the adopted reference frames were: 0.008 (0.024), 0.005 (0.018), 0.007 (0.017), 0.002 (0.015), and 0.014 (0.017), respectively.  With all of the data now on the same internal zero-point, the magnitude measurements and errors were combined for each filter via a weighted mean.  The weights were determined by the inverse variance returned by DAOPHOT.

The absolute calibration for each filter was determined by comparing the SDSS Stripe 82 magnitudes measured here against a reference set converted onto the natural \emph{DECam} system  following the procedure described in \citet{Calamida17}.  As a result, we only needed to calculate and apply a constant offset value for each band and could ignore color corrections. The accuracy of the calibration is $\approx$ 5\% for the bluer filters ($ug$) and $\approx$ 3\% for the redder filters ($ri$).

\begin{figure}
\includegraphics[width=\columnwidth]{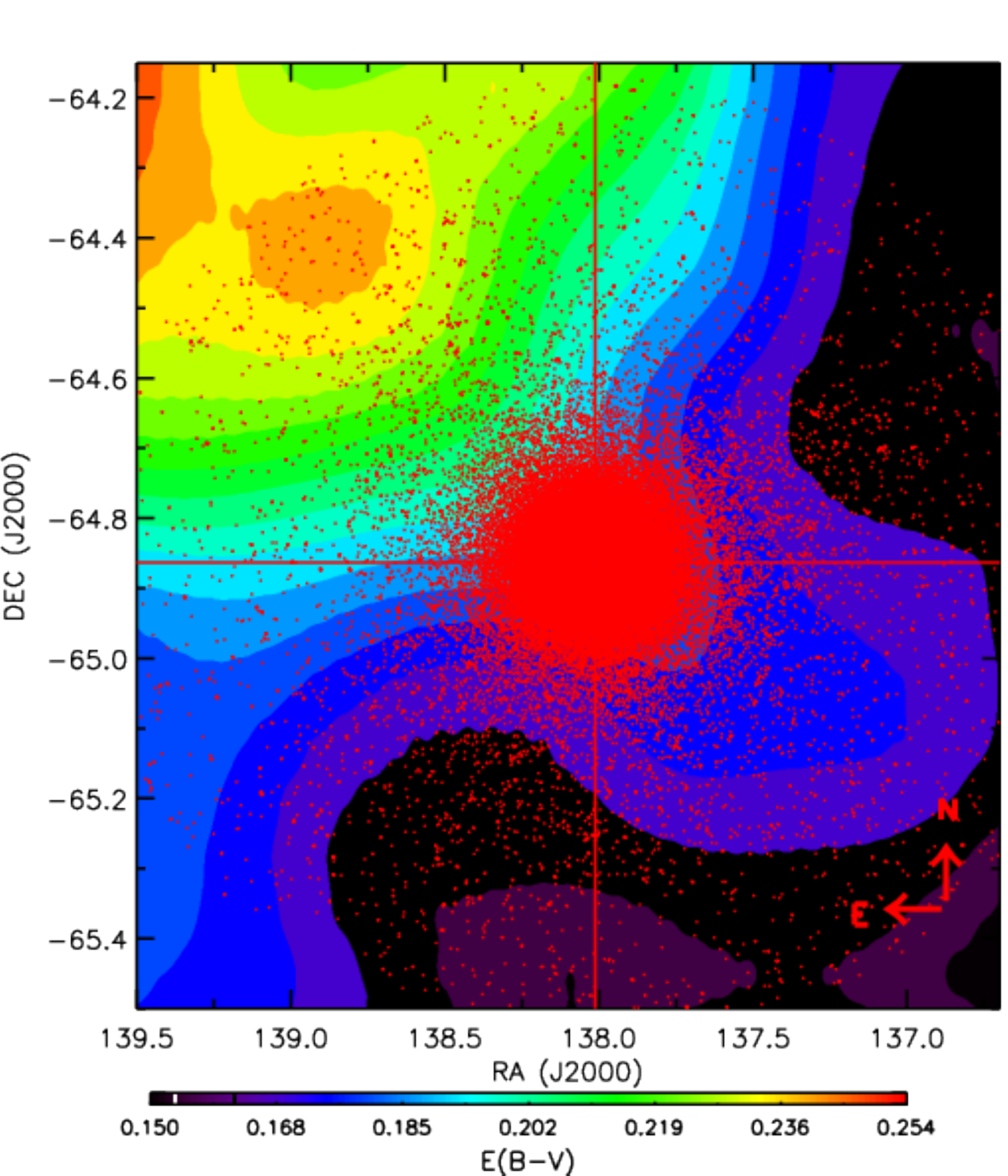}
\caption{Reddening color density map as derived from \citet{schlafly2011} for stars towards the observed region across NGC~2808. Stars identified as cluster members from \emph{DECam}   photometric catalog are over-plotted as red dots. The North and East direction are indicated with red arrows.}
\label{fig:red_map}
\end{figure}

Although NGC~2808 is mildly affected by reddening along the line-of-sight \citep[E(B-V) $\approx$ 0.2]{Bedin2000}, the differential reddening is relatively small across our FoV. Fig.~\ref{fig:red_map} shows a reddening map for the FoV towards NGC~2808 based on the extinction values provided by \citet{schlafly2011}. Cluster stars are overplotted as red dots in the plot. The figure shows that reddening is quite homogeneous for the central part of the cluster, while it may vary towards the outskirts. In particular, extinction seems slightly higher on the North--East quadrant of the FoV and lower in the South--West one. However, these regions are at and beyond the nominal tidal radius of NGC~2808 (22\arcmin) and only include very few cluster stars ($\approx$ 1 \%). Moreover, the extinction for the entire 1.5$\times$1.5 degree FoV is on average $E(B-V) \approx$ 0.18 mag with a dispersion of $\sigma =$ 0.03 mag. 
We also compared these reddening values with those from Gaia DR3 towards the FoV of NGC~2808 and obtained very similar results. Therefore, we did not apply any correction for differential reddening to our photometric catalog.  

\begin{figure*}
\includegraphics[width=0.55\textwidth,angle=90]{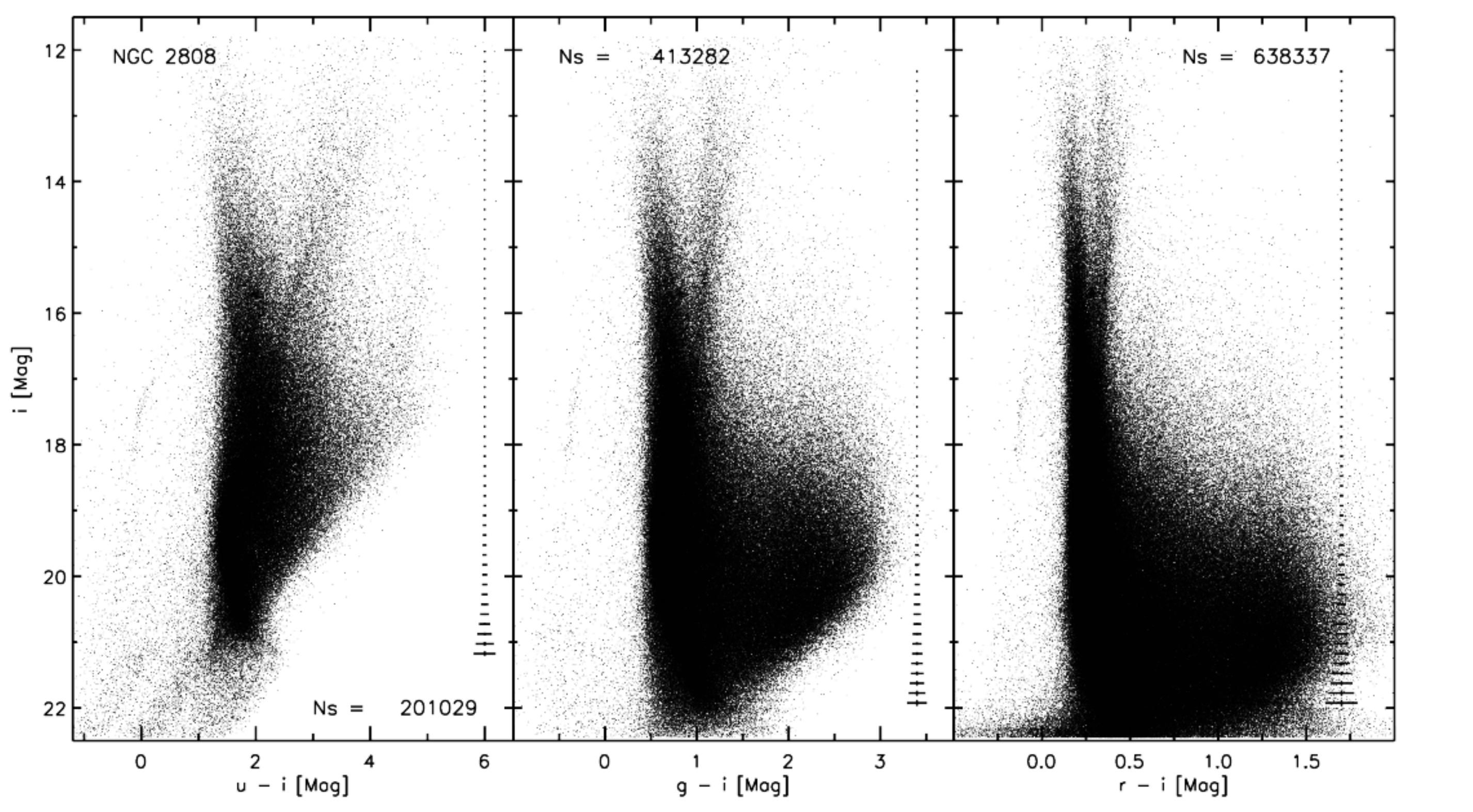}
\caption{$i$ versus $u-i$, $g-i$, and $r-i$ CMDs are shown for the full 
DECam field-of-view.  The cluster RGB, at 12.0 $\lesssim i \lesssim$ 16.0 and 1.4 $\lesssim g - i \lesssim$ 2.5 mag, and blue HB, at 16.0 $\lesssim i \lesssim$ 18.5 and $g - i \lesssim$ 0.5 mag, are clearly 
visible despite significant field star contamination. Error bars are shown. }
\label{fig:2808_cut}
\end{figure*}

The final \emph{DECam}   photometric catalog includes 1,990,974 objects measured in the FoV. The largest number of stars was detected in the reddest filter ($i$; $N=984,942$), and the fewest stars were measured in the $u$ filter ($N = 325,898$). The photometric catalog reaches a depth of $i \approx$ 21.5 mag with a signal-to-noise ratio ($S/N$) of $\approx$ 50. If we limit the photometry to observations including the $u$ filter, the depth is $i \approx$ 21 mag with $S/N \approx$ 70.

Fig.~\ref{fig:2808_cut} shows the $i,\ u-i$, the  $i,\ g-i$ and the $i,\ r-i$ color-magnitude diagrams (CMDs) for the entire sample of stars observed with \emph{DECam}   towards NGC~2808. The catalog was filtered by photometric accuracy, and $\approx$ 70\% of the best measured stars are plotted. Stars were also filtered by radial distance, to avoid the crowded regions of the cluster center, $r \ge$ 1.5\arcmin. 
The number of selected stars for each CMD is labeled in the figure.

\begin{figure*}
\includegraphics[width=0.55\textwidth,angle=90]{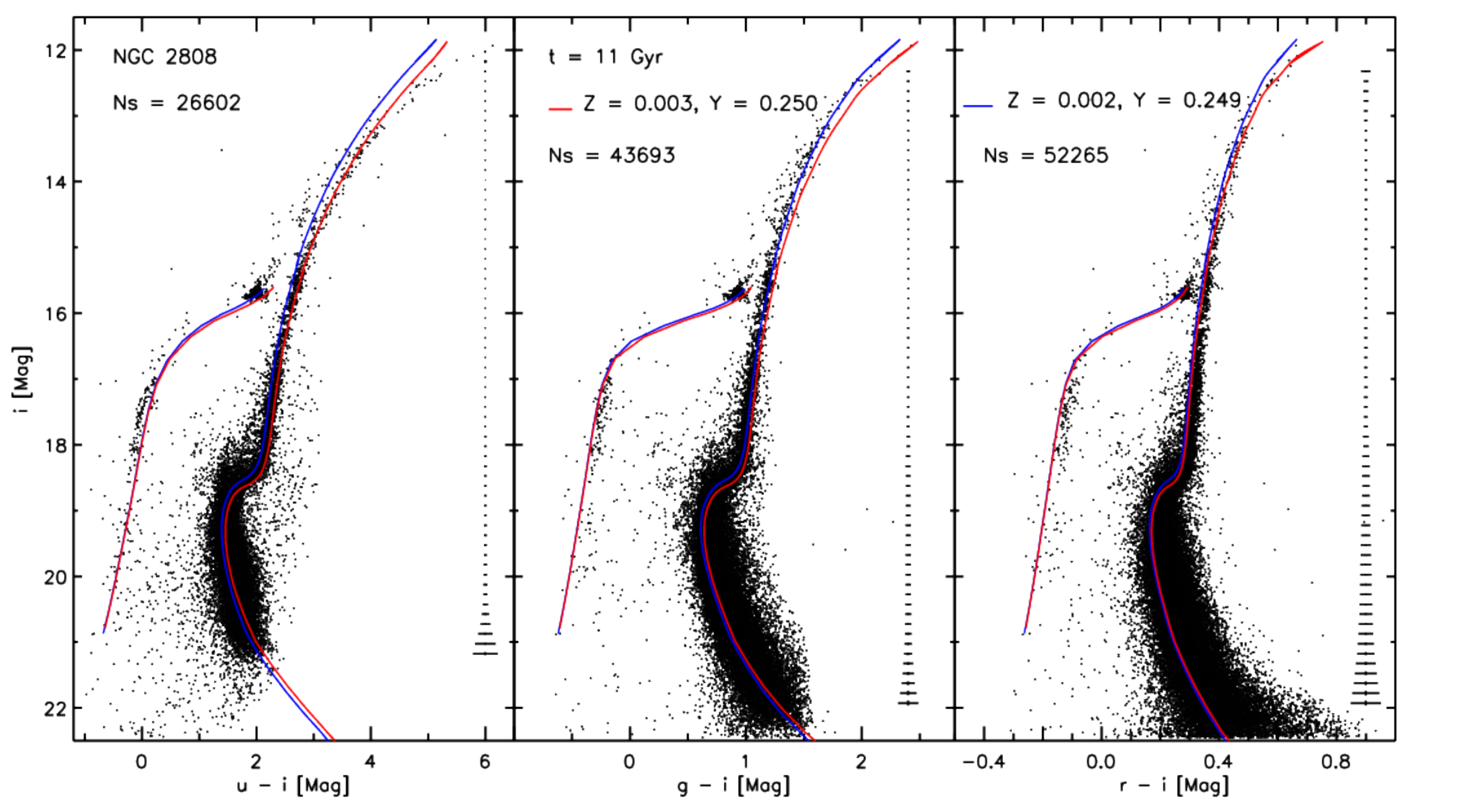}
\caption{Same CMDs to those shown in Fig.~\ref{fig:2808_cut} except
only candidate cluster members are shown.  The CMDs are compared
against BASTI isochrones and ZAHB models for different abundances and same age as labeled in the figure. A distance modulus of $\mu_0$ = 15.05 and a reddening of $E(B-V)$ = 0.18 mag were used. Error bars are shown. See text for more details.}
\label{fig:2808_iso_cut}
\end{figure*}

Fig.~\ref{fig:2808_cut} clearly shows that NGC~2808 CMDs are strongly contaminated by field stars. However, some cluster evolutionary sequences are detectable, such as the HB for 16 $\lesssim i \lesssim$ 18.5 and $g -i \lesssim$ 0.5 mag, and the RGB for 12.0 $\lesssim i \lesssim$ 16.0 and 1.4 $\lesssim g - i \lesssim$ 2.5 mag. On the other hand, the lower part of the RGB, the main-sequence turn-off (MSTO), and the lower MS are completely mixed with field stars. 
Gaia DR3 (Gaia collaboration et al.\ 2022) proper motion data for NGC~2808 are not complete in the more central cluster regions and they have a limiting magnitude of $G =$ 21, which is only 1 mag below the MSTO. Therefore, to separate the cluster and field components, we used the same approach devised by \citet{Calamida17, calamida2020}. Briefly, we took advantage of the $u$-band observations to create a color-color-magnitude plane, $r$ vs $g-i$ vs $u-r$, which better separates cluster and field stars due to their different metallicities and gravities. We utilized an iterative procedure to select 74,262 candidate NGC~2808 member stars with at least one measurement in the $i$ and $r$ filters. The final cleaned catalog has 36,826 candidate cluster members with at least one measurement in all filters, $ugri$, and the CMDs are shown in Fig.~\ref{fig:2808_iso_cut}.  

To verify the accuracy of our selection of cluster and field stars, we took advantage of Gaia DR3 data for the brighter portion of the photometric catalog. By matching using a radius of 0.5\arcsec we found 12,702 stars in common with a \emph{DECam}   measurement in all filters and proper motion measurement from Gaia. 
We then used the proper motion plane to estimate how many stars might have been misidentified by our method as cluster stars. Gaia proper motion for NGC~2808 is $\mu_{\alpha}$ = 0.994$\pm$0.024 and $\mu_{\delta}$ = 0.273$\pm$0.024 mas/yr \citep{vasiliev21} and we selected
as candidate cluster members stars with -4 $< \mu_{\alpha} <$ 5 and -4 $< \mu_{\delta} <$ 5 mas/yr. Of the 12,702 stars selected as NGC~2808 members with our color-color-magnitude method and in common with Gaia, 1,139 are field stars according to proper motions, i.e. $\sim$ 9\%. 
We repeated the same procedure for stars selected as field members from the color selection and less than 1\% are candidate cluster stars according to proper motions. 

The clean sample of NGC~2808 stars is shown on the $i,\ u-i$,  the $i,\ g-i$ and the $i,\ r-i$ CMDs of Fig.~\ref{fig:2808_iso_cut}. All the cluster sequences are clearly visible now, including the HB, divided into a RHB, clustering at $i \approx$ 15.8 mag, and a blue HB, extending down to $i \approx$ 20.5 mag. The MSTO is at $i \approx$ 19 mag and the RGB extends from its base at $i \approx$ 18 up to $\approx$ 12.5 mag. The RGB bump is also visible at $i \approx$ 15.5 mag.

In order to verify the accuracy of our photometric calibration, we compared the clean \emph{DECam}   CMD of NGC~2808 with models. We used two $\alpha$-enhanced BASTI\footnote{http://basti-iac.oa-abruzzo.inaf.it/index.html} isochrones for the same age, t $=$ 11 Gyr, two different metallicities, namely Z $=$ 0.002 (blue solid line) and 0.003 (red), and two Zero-Age Horizontal Branch (ZAHB) tracks \citep{pietrinferni2021}. 
These metallicity values, -1.2 $\le [Fe/H] \le$ -1.0, bracket 
the iron abundance estimate from \citet{Carretta15}.
As a distance modulus we used $\mu_0 =$ 15.05 mag and reddening $E(B-V) =$ 0.185 \citep{schlafly2011}. This reddening value was converted into extinction in the \emph{DECam}   filters by using the \citet{Cardelli89} reddening law and the available \emph{DECam}   filter throughputs\footnote{Information about the \emph{DECam}   filter throughputs can be found at: https://noirlab.edu/science/programs/ctio/filters/Dark-Energy-Camera.}. We obtained $A_{i}=$ 0.63$\times A_V$ and $E(u - i)=$ 2.65$\times E(B-V)$, $E(g - i)=$ 1.70$\times E(B-V)$, and $E(r - i)=$ 0.65$\times E(B-V)$.

Fig.~\ref{fig:2808_iso_cut} shows that the agreement between theory and observations is very good over the entire magnitude range in all the three CMDs. The two isochrones bracket the NGC~2808 RGB and closely fit the MSTO, while the ZAHB models reproduce the HB from the red HB down to the blue tail.

\section{Gaussian Mixture Models} \label{subsec:GMM}

\begin{figure*}
\includegraphics[width=0.5\textwidth]{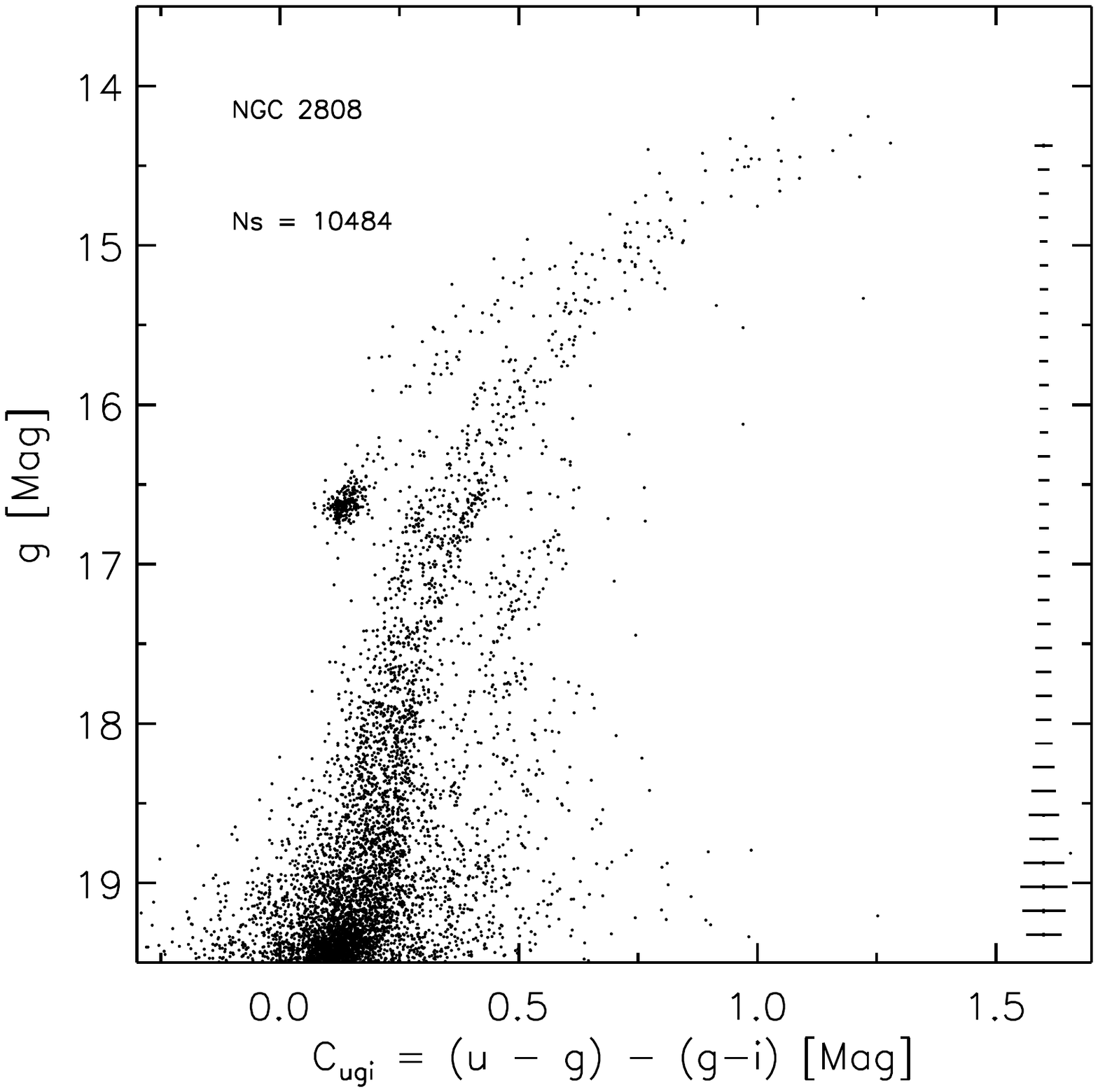}
\hspace{-1.05cm}
\includegraphics[width=0.5\textwidth]{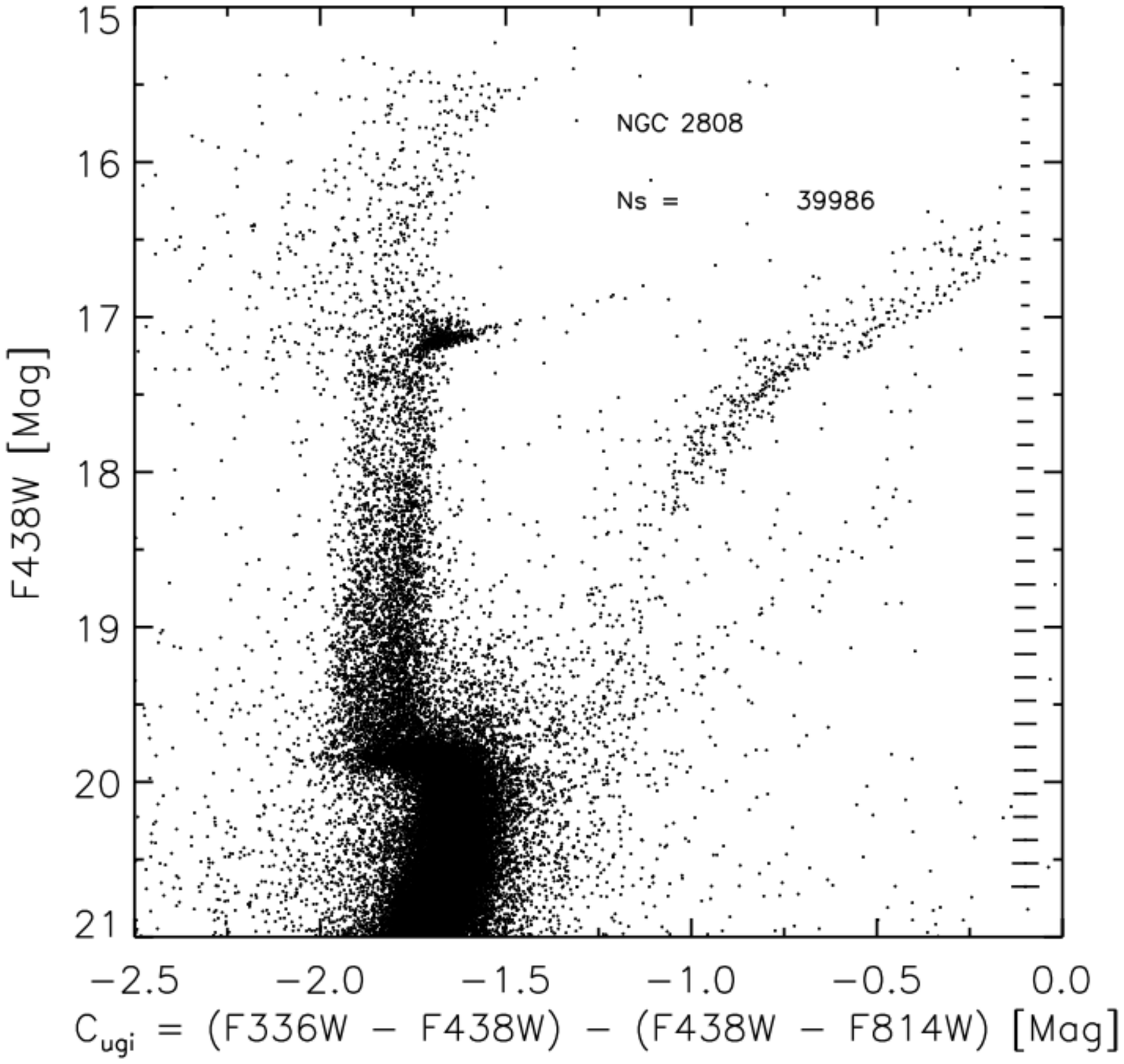}
\vspace{-1.5cm}
\caption{The $C_{ugi,DECam}$ CMD for NGC~2808 derived with \emph{DECam}   data is shown on the left panel while a similar $C_{ugi,HST}$ CMD derived with \emph{HST} data is shown on the right.  Both data sets support the existence of at least 2-3 distinct RGB sub-populations.  Note that the \emph{DECam}   data only
include stars outside 1.5$\arcmin$ from the cluster center while the \emph{HST} data trace stars inside $\sim$ 1.5$\arcmin$.}
\label{fig:cugi_plots}
\end{figure*}

We identify multiple stellar populations (MSPs) in NGC~2808 by fitting Gaussian mixture models (GMMs) to the distribution of the $C_{ugi}$ color index of RGB stars. Color indices involving blue or ultra-violet filters, e.g., $C_{ugi, \emph{DECam}  }$ = ($u-g$)$-$($g-i$), and the \emph{HST} equivalent $C_{ugi, HST}$ = ($F336W-F438W$)$-$($F438W-F814W$), are effective diagnostics for separating multiple stellar populations because they are sensitive to light element abundance variations. For example, star-to-star variations in $Na$ abundance are correlated with a spread in the $C_{ugi,DECam}$ color since $Na$-poor stars are bluer in $u-g$ and redder in $g-i$, similar to $U-B$ and $B-I$ \citep{Lardo11,Monelli13}. Sample $C_{ugi}$ CMDs using the \emph{DECam}   and \emph{HST} observations are shown in Fig.~\ref{fig:cugi_plots}, which highlights that both data sets produce multiple distinct RGB sequences.

\begin{figure*}
\includegraphics[width=\textwidth]{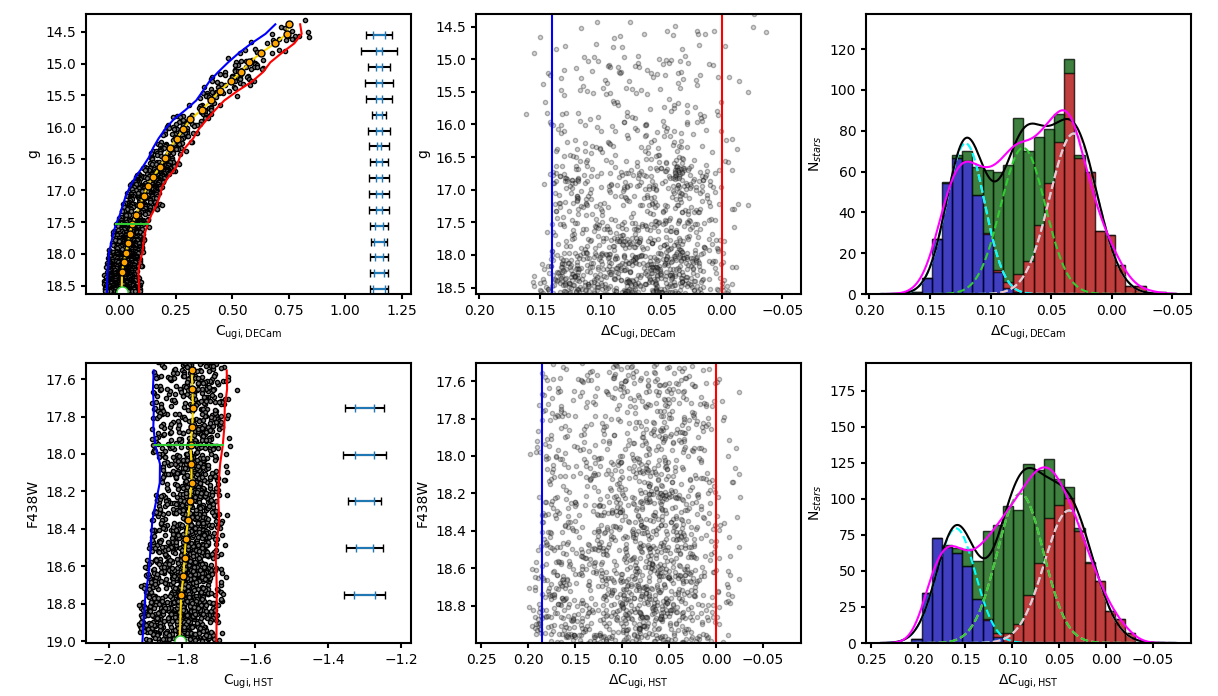}
\caption{Demonstration of the technique used to divide NGC~2808 RGB stars into individual populations using photometry from \emph{DECam} (top) and \emph{HST} (bottom). The left panels show the $g$ band (or equivalent $F438W$ for HST) versus $C_{ugi}$ CMDs. The orange line is the median fiducial ridge line and orange circles indicate the magnitude bin centers which are separated by 0.1 mag. The 4th and 96th quantile in $C_{ugi}$ as a function of apparent magnitude are indicated with the blue and red curves. The magnitude at which we measure the RGB width is shown as a green line which is one magnitude brighter than the faintest RGB star analyzed, which is indicated by the green open circle. The RMS spread in C$_{ugi}$ colors and the mean photometric error of stars along the RGB are shown as black and blue horizontal errorbars to the left. The center panels show the rectified RGBs, where the deviation from median RGB color is shown for each star as a function of magnitude. We collapse the rectified RGBs in the magnitude direction to reveal the individual sequences in $\Delta C_{ugi}$, shown in the right panels. The best-fitting GMMs are overlaid on the $\Delta C_{ugi}$ histograms: individual components are colored blue, green, and red, and the sum is the black curve. The Gaussian KDE of the distribution is shown as an magenta curve.} 
\label{fig:rgb_verticalized}
\end{figure*}

In order to construct the color distribution in the $C_{ugi}$ index, it is necessary to first rectify the RGB since the color distribution we are interested in should not include a contribution from the shape of the RGB. The color offset for each star is computed following the equation from \citet{Milone17_atlas}:

\begin{equation}
    \Delta C_{ugi} = W_{C_{ugi}}\frac{X_{\rm{fiducial\_R}} - X}{X_{\rm{fiducial\_R}}-X_{\rm{fiducial\_B}}},
\end{equation}

\noindent which is the RGB-width-scaled offset from the red edge of the RGB, where $X = C_{ugi}$, and the "fiducial R" and "fiducial B" correspond to the red and blue fiducial curves. $W_{C_{ugi}}$ is the width of the RGB measured one magnitude brighter than the faintest RGB star. We determined the red and blue fiducial curves by evenly dividing the RGB into equal-width magnitude bins, and in each bin computing the 4th and 96th percentiles in $C_{ugi}$ color. The rectification procedure is summarized schematically in the left and center panels of Fig.~\ref{fig:rgb_verticalized}.

\begin{figure}
\includegraphics[width=0.5\textwidth]{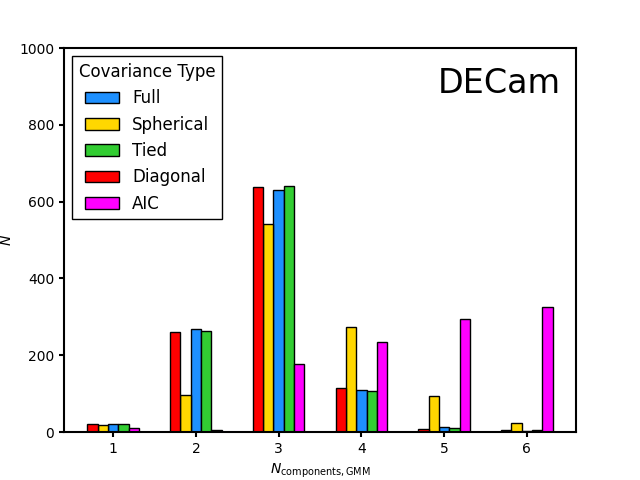}
\includegraphics[width=0.5\textwidth]{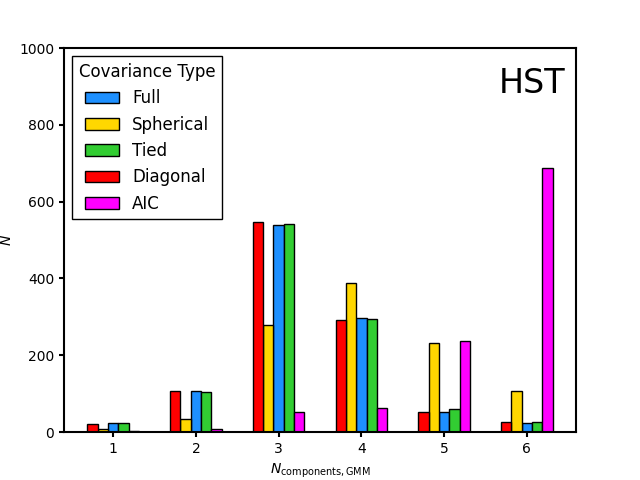}
\caption{The various histograms show the optimal number of components derived from GMM realizations of the data using the BIC with all components sharing the same covariance matrix ("tied"; green), each component having a separate diagonal covariance matrix ("diagonal"; red), each component having its own covariance matrix ("full"; blue), and each component having a single variance value ("spherical"; yellow).  The magenta histogram shows similar results but using the Akaike Information Criterion (AIC).  Most methods found that a three component fit was the optimal number.} 
\label{fig:Bayes}
\end{figure}

Stars inside and outside of r = 1.5$\arcmin$ were analyzed separately using \emph{HST} and \emph{DECam}   photometry, respectively. Inner cluster stars were selected to have $-2.5 < \rm{C_{ugi,HST}} < -1.6$ and $17.5 < \rm{F438W} < 19$. The outer cluster stars were selected to have $-0.25 < \rm{C_{ugi,DECam}} < 1.25$ and $14 < g < 18.6$. In both cases, we cut stars with photometric errors 3$\sigma$ above the median photometric error, and rejected stars with membership probabilities lower than 90\%.

We estimated the number of sequences along the RGB using a Monte Carlo (MC) approach where GMMs are repeatedly fit to the $\Delta C_{ugi}$ color distribution. At each iteration, we randomize the stellar magnitudes according to the photometric errors in each band. We also iterate over different magnitude bin widths ($0.05 < \Delta g, \Delta\rm{F438W} < 0.25$) and the number of sigma clips to the RGB ($1 < N_{\sigma} < 20$). The MC simulation resulted in $\sim$1000 realizations of the $\Delta C_{ugi}$ distribution. For each realization, the best of six GMMs (with $N_{\rm{components}}$ = 1--6) was determined using the Bayesian Information Criterion (BIC), which is similar to the $\chi^2$ goodness of fit statistic, but includes a term that penalizes for overfitting. The most frequently occurring value of $N_{\rm{components}}$ was taken as our estimate for the number of sequences along the RGB. This procedure was done separately for stars inside 1.5$\arcmin$ (using \emph{HST} photometry), and for stars outside 1.5$\arcmin$ (using \emph{DECam} photometry). Fig.~\ref{fig:Bayes} shows that for both data sets we find the three-component GMMs best fit the data.

Population tagging of RGB stars was done using component membership probabilities assigned to the stars according to the best-fitting GMM. The probability that the $i^{th}$ star is a member of the $j^{th}$ mixture component ($z_j$) was estimated using the measured value of $\Delta C_{ugi,i}$ as input to the probability mass function for the component: $p(z_j=1|{\bf{x}}_i=\Delta C_{ugi,i}) = {\rm{P}}_j(\Delta C_{ugi,i})$. The rightmost panels of Fig.~\ref{fig:rgb_verticalized} show the $\Delta C_{ugi,HST}$ distribution for RGB stars inside (bottom) and outside (top) r = 1.5$\arcmin$, with the best-fitting GMM overlaid. The \emph{HST} and \emph{DECam}   photometric decompositions were made using magnitude bins with a width of 0.1 mag. For the \emph{HST} data, we clipped stars in each bin with colors $> 3\sigma_{\rm{C_{ugi}}}$, while for the \emph{DECam}   data we clipped stars with colors $> 5\sigma_{\rm{C_{ugi}}}$.

\section{Literature Comparison} \label{sec:comparison}
\subsection{Stellar sub-population definitions} \label{subsec:pop_defs}
Although multiple chemically distinct groups have been found in NGC~2808, the nomenclature, separation of stars, and number of groups identified depends strongly on the data and analysis methods.  For example, \citet{Piotto07} identified three MSs with different helium abundances via \emph{HST} photometry and referred to these populations as "rMS", "mMS", and "bMS".  However, \citet{Latour19} used a combination of \emph{HST} photometry and MUSE/VLT (ESO) spectroscopy to identify four RGB populations (P1, P2, P3, and P4), while \citet{Hong21} used CN, CH, and Ca HK spectral indices to also find four RGB groups (G1, G2, G3, and G4).  Furthermore, \citet{Carretta15} used various light element abundance ratios to separate NGC~2808 stars into five groups (P1, P2, I1, I2, and E), and \citet{Milone15_n2808} used \emph{HST} "chromosome maps" to identify five slightly different populations (A, B, C, D, and E).

More recently, \citet{valle2022} used robust statistical methods to identify the different stellar populations in NGC~2808 by combining the high-resolution spectroscopy of \citet{Carretta15} and the low-resolution spectroscopy of \citet{Hong21}, and found only two groups along the cluster RGB, further complicating matters.

\begin{figure*}
\includegraphics[width=\textwidth]{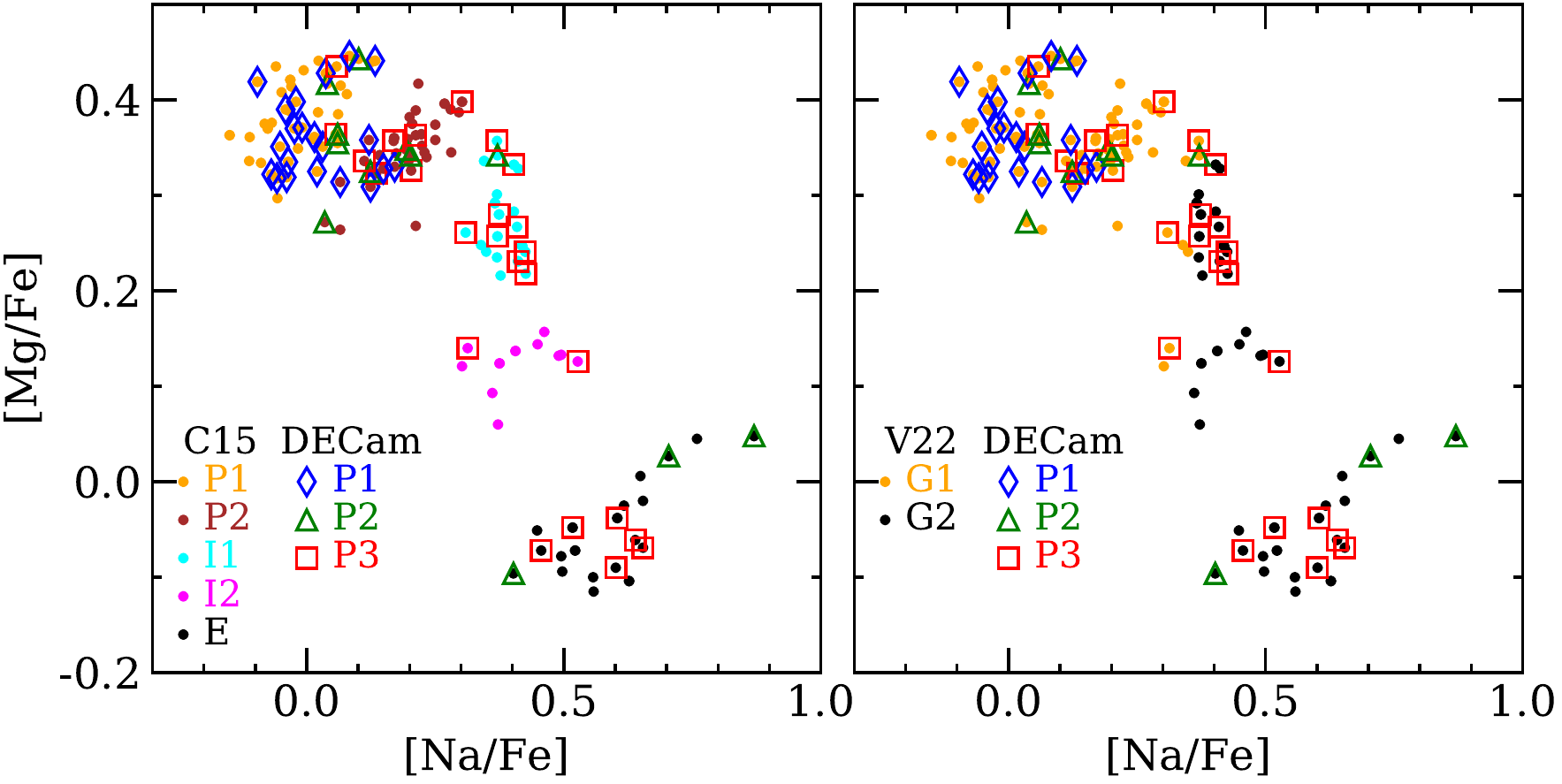}
\caption{The left panel shows the [Mg/Fe] versus [Na/Fe] abundances from \citet{Carretta15} for NGC~2808 RGB stars.  Each of the five populations identified by \citet{Carretta15} is shown as a different colored filled circle.  Cross-matched populations from our \emph{DECam} $C_{ugi,DECam}$ decomposition are shown as different colored large open symbols.  In general, we find that both studies cleanly separate primordial (P1) stars from those that formed from gas processed at higher temperatures.  The $C_{ugi,DECam}$ index generally finds that P3 stars have higher [Na/Fe] and lower [Mg/Fe] than those in the P2 group, but both populations are somewhat mixed.  The mixing of P2 and P3 stars is likely because the \citet{Carretta15} sample mostly consists of bright RGB stars where the $C_{ugi,DECam}$ separation is small.  Similarly, the right panel shows the same data but with the \citet{Carretta15} sample separated into only two groups, based on the reanalysis by \citet{valle2022}.}
\label{fig:carretta_comp}
\end{figure*}

\begin{figure}
\includegraphics[width=0.45\textwidth,height=0.7\textheight]{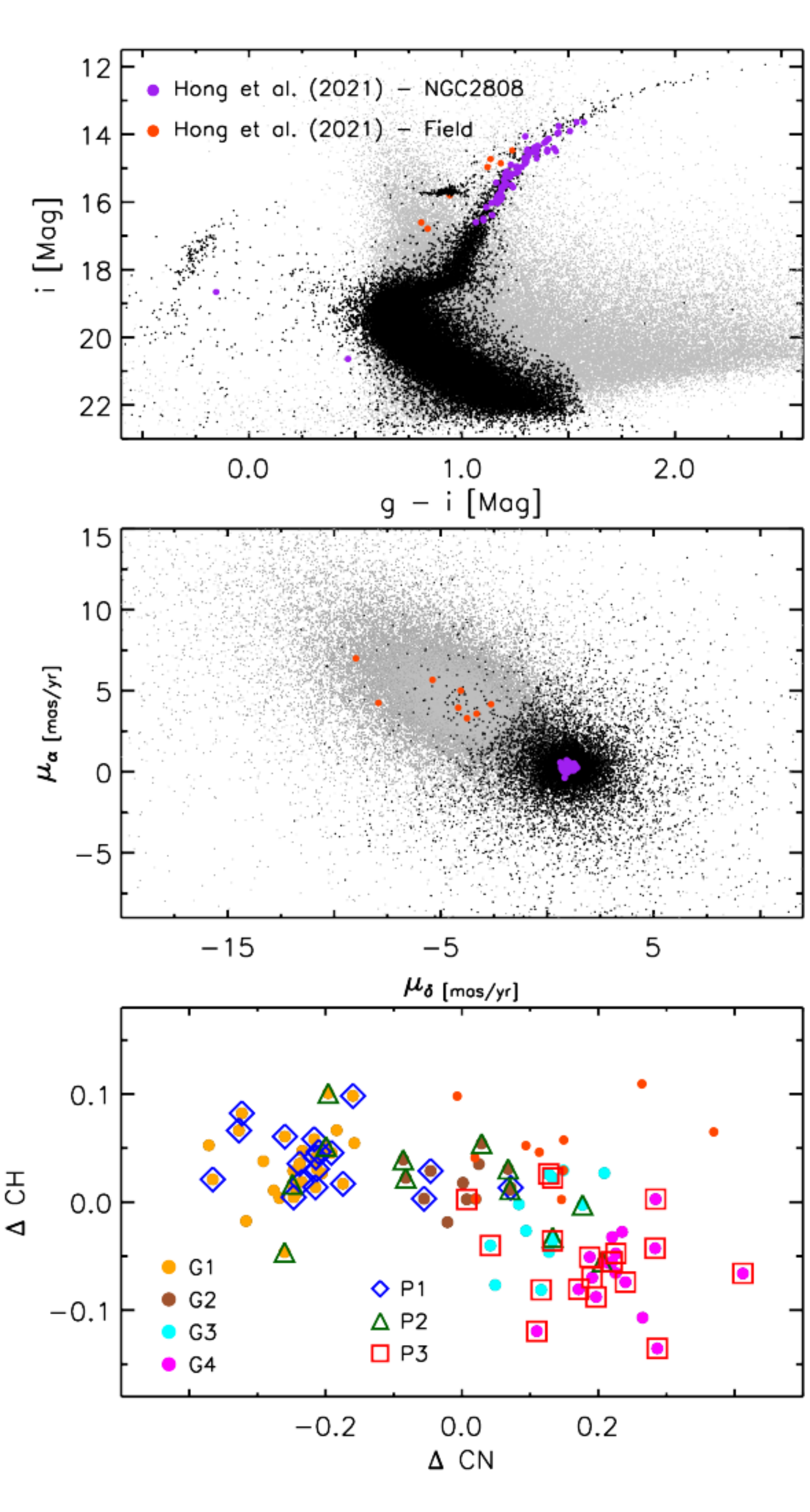}
\caption{Top: \emph{DECam}   $i,\ g-i$ CMD for NGC~2808 candidate member stars (black dots) and for candidate field stars (grey) according to the color-color-magnitude selection. The 82 stars in common with the spectroscopic study of Hong et al. (2021) are overplotted as purple (NGC~2808 members) and orange dots (field). Two of the candidate cluster members are not on the RGB (see text for more details). Middle: same stars plotted on the Gaia proper motion plane. Bottom: the 82 stars with spectroscopy are plotted on the $\Delta CN$ vs $\Delta CH$ plane: the four groups identified by Hong et al. are marked with different colors and labeled as $G1, G2, G3,$ and $G4$. The stars for which we derive the $C_{ugi}$ index and separate in three groups are marked with different symbols and labeled as $P1, P2,$ and $P3$. The 10 stars from Hong et al. that are classified as field members by our color-color-magnitude selection and Gaia proper motions are marked as orange filled circles.}
\vspace{-0.2cm}
\label{fig:hong}
\end{figure}

Table 1 of \citet{Dantona16} provides an approximate mapping between the \citet{Carretta15} and \citet{Milone15_n2808} groups, but the connection to similar nomenclature in other works  is not straightforward.  

Our GMM grouping algorithm identified three populations as the optimal number, regardless of whether the $C_{ugi,DECam}$ (ground-based) or $C_{ugi,HST}$ (space-based) data were used.  We label these three populations as the P1, P2, and P3 groups, which correspond to stars having "primordial", "intermediate", and "extreme" chemical compositions.  The P1, P2, and P3 populations constitute 31$\%$, 38$\%$, and 31$\%$ of our total RGB sample (3060 stars), respectively.  We compared our $C_{ugi,DECam}$ designations against those of \citet{Carretta15}, \citet{valle2022}, and \citet{Hong21} in Figs.~\ref{fig:carretta_comp}-\ref{fig:hong}, and correlated our $C_{ugi,HST}$ populations with those found in \citet{Latour19} in Fig.~\ref{fig:hst_pops}.

\subsection{Stellar sub-population matching} \label{sec:matching}
First, comparing our $C_{ugi,DECam}$ populations against those of \citet{Carretta15} in Fig.~\ref{fig:carretta_comp}, we found that the strongest correlation is between the "primordial" (P1) groups of both studies.  For example, using the stars in common between the two studies we found that 76$\%$ (16/21) of our P1 stars overlap with the P1$_{C15}$ population, the remaining 24$\%$ align with the adjacent P2$_{C15}$ group, and none align with the I1$_{C15}$, P2$_{C15}$, or E$_{C15}$ groups.  However, the correlations become more complicated for the enriched populations.  Our P2 group, which is more chemically enhanced (lower $[O/Fe]$; higher $[Na/Fe]$) than the P1 group, mildly overlaps with the P1$_{C15}$, P2$_{C15}$, I1$_{C15}$, and E$_{C15}$ groups.  Similarly, our P3 group overlaps with 25 stars in the \citet{Carretta15} sample, and these stars are distributed as 8$\%$ (2), 24$\%$ (6), 36$\%$ (9), 8$\%$ (2), and 24$\%$ (6) in the P1$_{C15}$, P2$_{C15}$, I1$_{C15}$, I2$_{C15}$, and E$_{C15}$ populations, respectively.  Therefore, we can align our P1 group with the P1$_{C15}$ population with high confidence, and consider our P2$+$P3 groups to be a combination of the P2$_{C15}$, I1$_{C15}$, I2$_{C15}$, and E$_{C15}$ populations from \citet{Carretta15}.  We note that the poor correlation between $C_{ugi,DECam}$ and the \citet{Carretta15} designations, particularly for more enriched stars, is due to the latter work targeting primarily cool, bright RGB stars.  Fig.~\ref{fig:cugi_plots} shows that the color separation in $C_{ugi,DECam}$ is more narrow for stars significantly brighter than the RGB-bump, and also that AGB confusion increases for bright giants.

The right panel of Fig.~\ref{fig:carretta_comp} shows that the correlations are somewhat stronger when adopting the 2 population model from \citet{valle2022}.  In this scenario, the Group 1 population from \citet{valle2022} contains 100$\%$ of our overlapping P1 population, 75$\%$ of our P2 stars, and 44$\%$ of our P3 group while their Group 2 population is almost entirely composed (83$\%$; 15/18) of our most chemically enriched P3 stars.  Combining the information from both panels of Fig.~\ref{fig:carretta_comp} suggests that the $C_{ugi,DECam}$ color is highly sensitive for separating primordial and "second generation" stars from each other, but that more nuanced separations with only these filters are difficult when analyzing only bright RGB stars.  Therefore, a comparison between our $C_{ugi,DECam}$ population separation and that of \citet{Hong21}, which observed warmer stars, may provide more information about how the $C_{ugi,DECam}$ color separation correlates with populations identified via spectroscopy.

\begin{figure*}
\includegraphics[width=\textwidth]{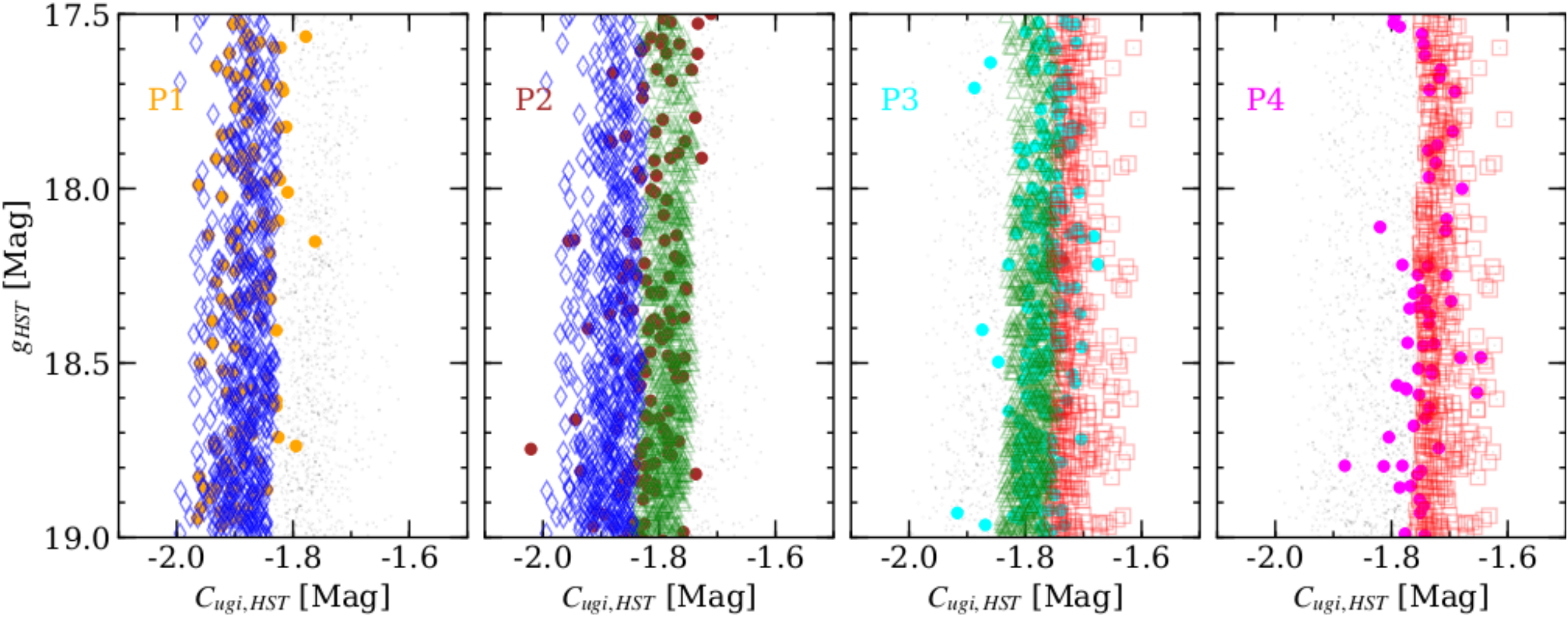}
\caption{$C_{ugi,HST}$ CMDs are shown with the four populations from \citet{Latour19}, which were identified using chromosome maps that include the $F275W$ filter, indicated as large orange, brown, cyan, and magenta circles.  The open symbols represent the \emph{HST} equivalent P1 (blue), P2 (green), and P3 (red) populations identified in Fig.~\ref{fig:carretta_comp} for our \emph{DECam}   data and use the same color/symbol shape scheme.  The small, light grey circles illustrate the full width of the $C_{ugi,HST}$ color range.  The comparison shows that the P1 group from \citet{Latour19} is strongly mapped to our P1 group, their P2 group is relatively well-aligned with our P2 group, and their P3/P4 groups are a mixture of our P2, but mostly P3, groups.  This figure highlights the strong correlation between the $C_{ugi,HST}$ index and similar indices that use $F275W$.}
\label{fig:hst_pops}
\end{figure*}

Fig.~\ref{fig:hong} plots 82 stars in common between the present study and \citet{Hong21} in the $\Delta CH$ vs $\Delta CN$ plane.  We found a good correspondence between the $C_{ugi,DECam}$ photometry and spectroscopic populations, and a significantly reduced scatter compared to the results shown in Fig.~\ref{fig:carretta_comp} for the brighter sample from \citet{Carretta15}.  We found that 83$\%$ (15/18) of the stars identified as belonging to the P1 group from $C_{ugi,DECam}$ photometry align with the G1 group from \citet{Hong21} with the remaining 3 stars overlapping with the G2 group.  Similarly, 94$\%$ (17/18) of our P3 stars align with the G3 and G4 populations from \citet{Hong21}.  Our P2 population has the largest overlap (42$\%$; 5/12 stars) with the G2 group, but also has a handful of stars in the G1 (33$\%$; 4/12), G3 (17$\%$; 2/12), and G4 (8$\%$; 1/12) groups.  Therefore, we can associate our P1 and P2 populations with those of the G1 and G2 groups from \citet{Hong21}, and also find that their G3 and G4 groups combine to match our P3 population.  

The top and middle panels of Fig.~\ref{fig:hong} also validate our membership selection procedure for NGC~2808.  In these panels, the black and grey circles indicate cluster members and field stars separated using our color-color-magnitude selection, respectively, while the purple and red symbols show stars from \citet{Hong21} that are cluster members and field stars using the same criteria.  The purple circles are located along NGC~2808's RGB sequence, as expected, and the red dots overlap with field stars.  In the proper motion plane (middle panel), the purple circles overlap with the cluster stars while the red circles are clearly offset with the field stars.  When the handful of field stars are identified in the $\Delta$CH-$\Delta$CN plane (bottom panel), it becomes clear that these stars cause an artificial enhancement in the scatter of the nominal $CH-CN$ anti-correlation.  Two of the targets from \citet{Hong21} are members according to their proper motions, but are clearly offset from the cluster sequence.  We suspect that these two stars were misidentified in the cross-match and have been removed from the analysis.

Finally, Fig.~\ref{fig:hst_pops} shows that we found a similar correlation between the $C_{ugi,HST}$ populations identified here and those found by \citet{Latour19}, which used the same data but included the $F275W$ filter.  Our P1 group is strongly correlated with the P1$_{L19}$ population, and in general our P2 group aligns well with the P2$_{L19}$ stars.  However, we found some mild overlap between our P2 group and the P3$_{L19}$ stars.  Similarly, our P3 population matches a combination of the P3$_{L19}$ and P4$_{L19}$ groups.  Therefore, we found a similar result when comparing with \citet{Latour19} as with \citet{Hong21} where our P1 group aligns with the P1$_{L19}$ stars, our P2 group matches the P2$_{L19}$ group, and our P3 population aligns with a combination of the P3$_{L19}$ and P4$_{L19}$ groups. 

A summary of the mapping between our three \emph{HST} and \emph{DECam}   populations and the various groups identified by \citet{Carretta15}, \citet{Latour19}, \citet{Hong21}, and \citet{valle2022} is provided in Table~\ref{tab:table3}.  In the next section, we will examine the radial profiles of the groups listed in this table and investigate differences in their distributions.

\begin{table*}[!htbp]
    \centering
    \caption{Population Correspondence}
    \begin{tabular}{llll}
    \hline
    Reference   &   P1$_{ours}$  &   P2$_{ours}$ &   P3$_{ours}$ \\
    \hline
    \citet{Carretta15}  &   P1$_{C15}$  &   (P2 + I1 + I2 + E)$_{C15}$  &   (P2 + I1 + I2 + E)$_{C15}$ \\
    \citet{valle2022}   &   Group 1 + Group 2   &   Group 1 + Group 2   &   Group 2 \\
    \citet{Hong21}  &   G1  &   G2  &   G3 + G4 \\
    \citet{Latour19}    &   P1$_{L19}$  &  P2$_{L19}$   &   (P3 + P4)$_{L19}$ \\
    \hline
    \end{tabular}
    \label{tab:table3}
\end{table*}

\section{Radial Distributions}\label{subsec:radial}
As noted previously, \citet{Carretta15} used high-resolution spectroscopy to divide a sample of 140 NGC~2808 RGB stars into five stellar populations (labeled P1$_{C15}$, P2$_{C15}$, I1$_{C15}$, I2$_{C15}$, and E$_{C15}$ in Section \ref{subsec:pop_defs}).  The data were further partitioned into three main groups: primordial (P), intermediate (I), and extreme (E), which aligned with the dominant populations identified in \citet{Carretta09_gir}.  When the radial distributions of the three main groups were analyzed, the author found that the combined I + E populations were more centrally concentrated than the P group, at least for distances between $\sim$ 1$\arcmin$ and 5$\arcmin$ from the cluster center, which is nominally in agreement with the cluster formation model described in \citet{Dercole08}.  However, \citet{Carretta15} noted that their sample size was too small to draw any firm conclusions regarding radial distance variations between sub-populations.

\citet{simioni2016} used WFC3 + ACS imaging on \emph{HST} to photometrically separate the three dominant MS groups, which likely trace populations with different helium abundances \citep{Piotto07}.  The authors found that the blue (most extreme/He-rich) MS is the most centrally concentrated while the middle MS stars are more concentrated than the red MS but less concentrated than the blue MS.  Related to this work, \citet{bellini20152015} measured proper motions for a sample of NGC~2808 MS stars using ACS on \emph{HST} and found no differences in velocity dispersion for the various MS populations as a function of radial distance.  However, their primordial MS groups (B and C) are nearly isotropic while the more enriched populations (D and E) are radially anisotropic; these most enriched stars also have smaller tangential velocity dispersions.

\citet{bellini20152015} showed that the strongest deviations from isotropy are in the outer parts of the cluster (for distances larger than $\approx$ 2 r$_{h}$).  Furthermore, the authors provide a simulation that demonstrates how enriched stars that are initially more centrally concentrated disperse with time on preferentially radial orbits \citep[see also][]{Mastrobuono13,Henault15,Mastrobuono16}.  As a result, many enriched stars that are now in the outer parts of the cluster may have initially formed in the core and dispersed outward, and thus a cluster's various sub-populations may have different kinematic and radial density profiles.  As long as a cluster's dynamical evolution is not too advanced, kinematic and radial density differences between primoridal and enriched stars may still be observable after a Hubble time \citep[e.g.,][]{Vesperini13,Mastrobuono16,Vesperini21,Tiongco22}.


NGC~2808 has a multi-modal HB, as shown by different ground- and space-based photometric investigations, with a RHB and a blue tail divided in three groups \citep{sosin1997, Bedin2000, castellani2006, iannicola2009}.
The origin of this multi-modal HB has been attributed to different "second" parameters, such as age, mass loss along the RGB due to rotation or binarity, the "hot-flasher" scenario, and/or to helium enrichment \citep{dcruz1996,Sweigart1998, catelan1998, brown2001, moehler2004, Dantona05, Lee05, castellani2006}.

\citet{walker1999} and \citet{Bedin2000} used ground-based photometry of NGC~2808 to study the radial distribution of the different groups of HB stars and found no gradient across most (4-6\arcmin) of the cluster body (note that NGC~2808 tidal radius is $r_t \approx$ 22\arcmin, see Table~\ref{table:1}).
However, a subsequent investigation based on UV \emph{HST} photometry by \citet{castellani2006} found a radial trend in the ratio of HB stars to the number of RGB stars brighter than the Zero-Age-Horizontal-Branch (ZAHB) luminosity level, the so-called $R$ parameter. In particular, this ratio increases from the expected value of $R \approx$ 1.4 \citep{cassisi2003, zoccali2000} in the cluster center to $\approx$ 1.7 at 2\arcmin~ distance. They propose different hypothesis for the extended distribution of the HB stars, such as high intensity mass loss along the RGB producing "blue hook" HB stars through the "hot-flasher" scenario, and/or a dynamical origin. Note that their \emph{HST} data set do not cover radial distances larger than $\approx$ 2\arcmin.

\citet{sohn1998} found a color gradient when analyzing ground-based photometric observations of NGC~2808, with the central regions ($r <$ 70\arcsec) being redder than the outskirt. As an explanation, they proposed an excess of RGB stars in the cluster core. On the other hand, \citet{sandquist2007} showed that NGC~2808 has a paucity of bright RGB stars compared to model predictions, possibly confirming a larger mass loss during this evolutionary phase, with a delayed or missed helium flash.

Furthermore, \citet{iannicola2009} confirmed a flat radial distribution of HB stars in NGC~2808, as previously found by \citet{walker1999} and \citet{Bedin2000}, based on \emph{HST} and ground-based photometry of the cluster. They also found that the $R$ parameter increases towards the outskirt of the cluster, and they explain this with a decrease of bright RGBs in these regions. However, their result might be hampered by the contamination of field stars that could artificially increase the number of RHB stars.

All the aforementioned results on the spatial distribution of the different RGB sub-populations and multiple HBs in NGC~2808 are based on heterogeneous photometric catalogs limited to radial distances less than $\approx$ 4\arcmin. 
Our deep and precise combined \emph{DECam} + \emph{HST} photometric catalog will now allow us to perform a thorough analysis of the radial distribution of the three different stellar sub-populations we identified on the RGB and the multiple HBs across the entire extent of NGC~2808.

\subsection{RGB stars}
The RGB radial distribution investigation executed in the present work includes $>$3000 high membership probability stars brighter than $g_{DECam}$ = 18.5 and $F438W_{HST}$ = 19.0 mag ranging from near the cluster core to about the tidal radius.  Since Section \ref{sec:matching} confirmed that the $C_{ugi}$ pseudo-color is strongly correlated with RGB light element composition, we can use this information to analyze the different radial density profiles of the P1, P2, and P3 populations.

\begin{figure*}
\includegraphics[width=\textwidth]{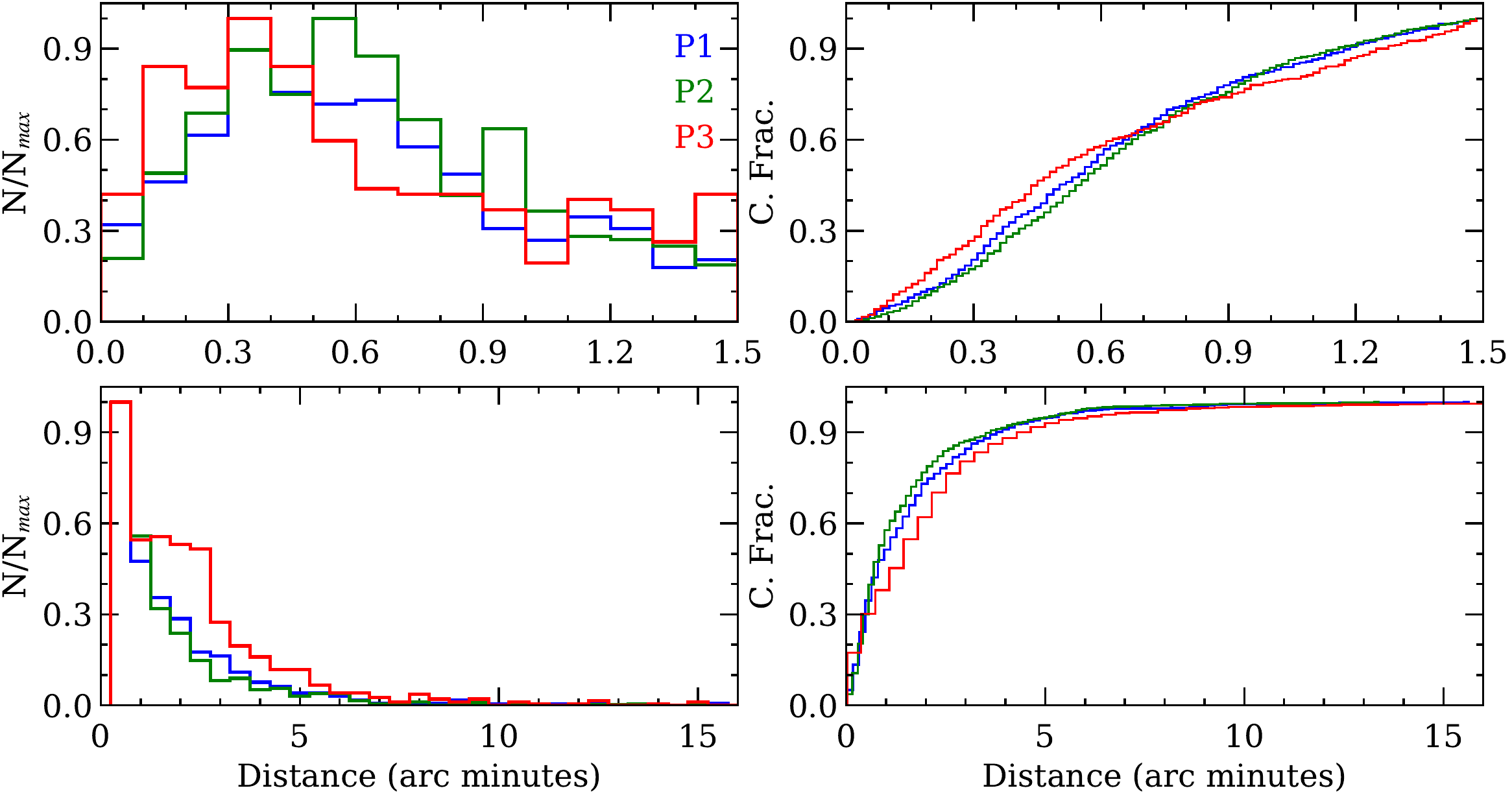}
\caption{\emph{Top left:} normalized radial distributions are shown for the P1 (blue), P2 (green), and P3 (red) populations inside 1.5$\arcmin$ from the cluster center.  The data are binned in 0.1$\arcmin$ increments, and are normalized relative to the most populated bin for each group.  \emph{Bottom left:} a similar plot using 0.5$\arcmin$ bins but extending out to about 70$\%$ of the tidal radius.  \emph{Top right:} cumulative radial distributions are shown for the three NGC~2808 populations using the same colors as the left panels.  \emph{Bottom right:} a similar cumulative distribution extending out to the tidal radius.  These panels show that the P3 group is centrally concentrated inside about 1 half-light radius (0.8$\arcmin$), and then becomes more dispersed in the outer parts of the cluster.  Similarly, the P2 group is the most centrally concentrated between about 1 and 5 half-light radii.}
\label{fig:radial_dist}
\end{figure*}

Fig.~\ref{fig:radial_dist} summarizes the different radial distribution profiles for the three main sub-populations identified from the GMM procedure outlined in Section 3, with the \emph{HST} and \emph{DECam} data analyzed separately.  The left panels of Fig.~\ref{fig:radial_dist} show the number of stars for a given sub-population within each radial bin but normalized to the bin of maximum height for each group.  Similarly, the right panels show the cumulative distributions of the same data.  These panels highlight a few interesting trends.  First, the P3 stars (most $Na$-rich according to the comparison with spectroscopy) appear to be the most centrally concentrated inside $\approx$ 0.7$\arcmin$ ($\approx$ 0.88r$_{h}$); however, the P3 trend flattens out substantially at larger radii.  For distances ranging from $\approx$ 1-5\arcmin~ from the cluster core, the P2 (intermediate enrichment) stars are more centrally concentrated than both the P3 and P1 (primordial) populations.  The flatter P3 radial density distribution extends out to at least 10 half-mass radii, but the number of stars in each sub-population becomes too small to differentiate trends between the three groups at larger distances.

\begin{figure}
\includegraphics[width=\columnwidth]{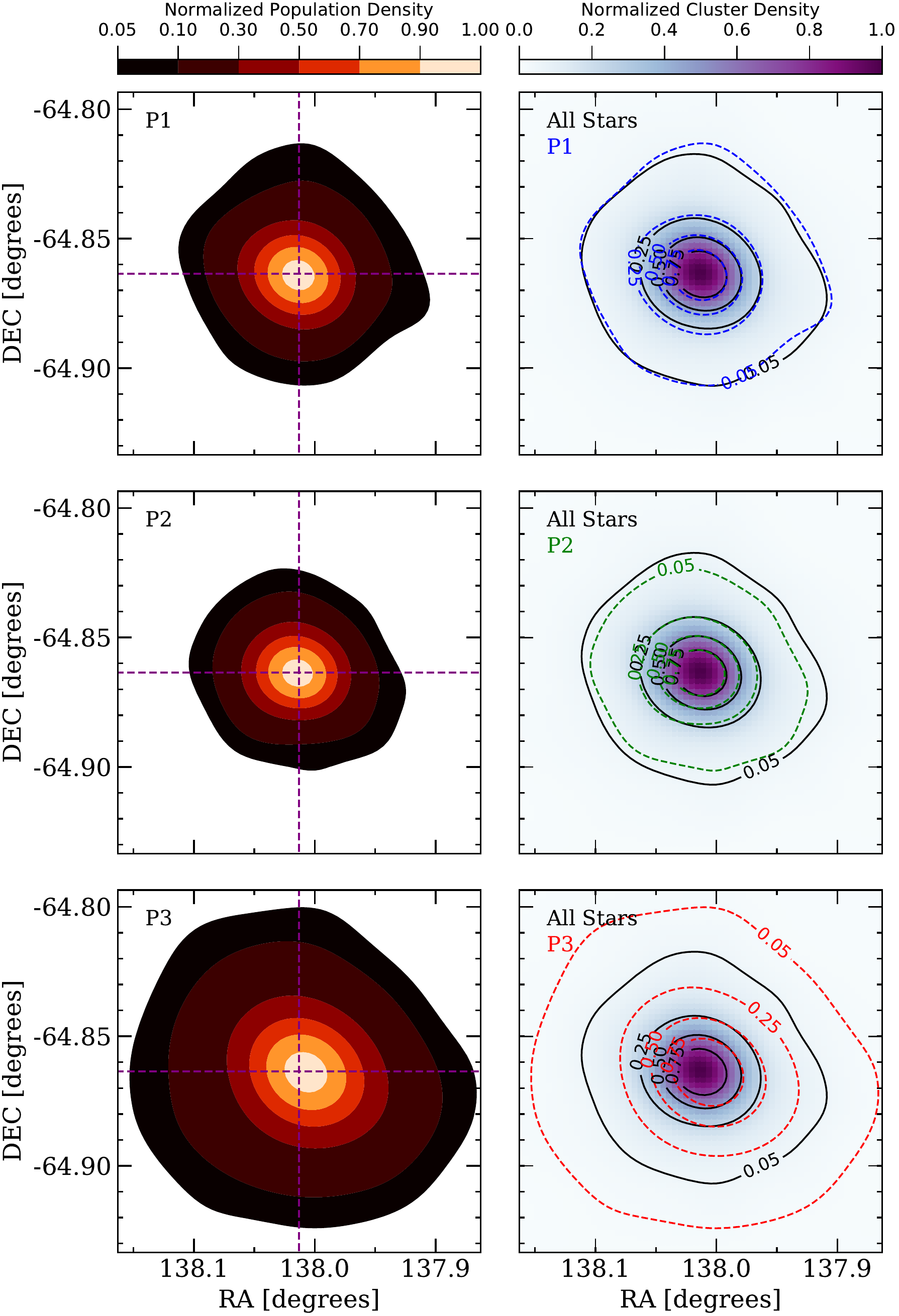}
\caption{\emph{Left:} normalized isodensity contours are shown for the P1 (top), P2 (middle), and P3 (bottom) populations.  The data are normalized such that the illustrated fraction is relative to the highest density bin for each population.  The intersection of the dashed purple lines highlights the adopted cluster center.  \emph{Right:} the color scale map shows a similar normalized density distribution to the left panels, but using all stars in the cluster.  The black contours are the same in each panel and illustrate the density distributions of the entire cluster.  The blue, green, and red contours are the same relative density levels for the P1, P2, and P3 populations, respectively.  The panels indicate that the P3 group is off-center from the P1 + P2 populations, and also show that the P3 stars are more broadly distributed.  The bottom right panel also indicates that the P3 stars may be aligned with a slightly different position angle than the rest of the cluster.}
\label{fig:cluster_iso}
\end{figure}

Fig.~\ref{fig:cluster_iso} compares normalized isodensity contours for the P1, P2, and P3 sub-populations over the full range of radial distances considered here.  The panels again highlight that the P2 sub-population is more centrally concentrated than the P1, and especially P3, groups when extending to distances far from the cluster center.  Similarly, Fig.~\ref{fig:cluster_iso} shows that the 50$\%$ contour level of the P3 group covers an area that is nearly 2$\times$ larger than either the P1 or P2 sub-populations.  The bottom right panel of Fig.~\ref{fig:cluster_iso} also shows mild evidence that the P3 population may have a slightly different position angle that is rotated northward compared to the P1 and P2 groups.  However, one of the most intriguing results from Fig.~\ref{fig:cluster_iso} is that the P3 sub-population centroid appears offset relative to the nominal cluster center, and also relative to the P1 and P2 centroids.  A Monte Carlo simulation with 10,000 resamplings of the population designations, based on the probabilities assigned in Section 4.2 to each star, indicates that the P3 group centroid is offset from the P1 and P2 population centers by $\approx$ 6.0-6.5$\arcsec$ (0.14r$_{h}$), with most of the shift coming from the right ascension coordinate.

\subsection{HB stars}\label{sec:hbrad}
We used here our \emph{DECam} + \emph{HST} deep and precise photometry, which covers more that the entire tidal extent of NGC~2808, to investigate the radial distribution of the HB stars.
As a first step, \emph{HST} photometry in the $F438W, F606W,$ and $F814W$ filters was converted into the \emph{DECam} $g, r,$ and $i$ filters. A sample of bright and well-measured stars in common between the two data sets was selected and the following color transformations were derived:

\begin{equation}
    g = F438W -0.05 - 0.27 \times (F438W - F814W)
\end{equation}

\begin{equation}
    r = F606W + 0.11 - 0.29 \times (F606W - F814W)
\end{equation}

\begin{equation}
    i = F814W + 0.27 + 0.14 \times (F606W - F814W)
\end{equation}

\begin{figure*}
\includegraphics[width=0.55\textwidth, angle=90]{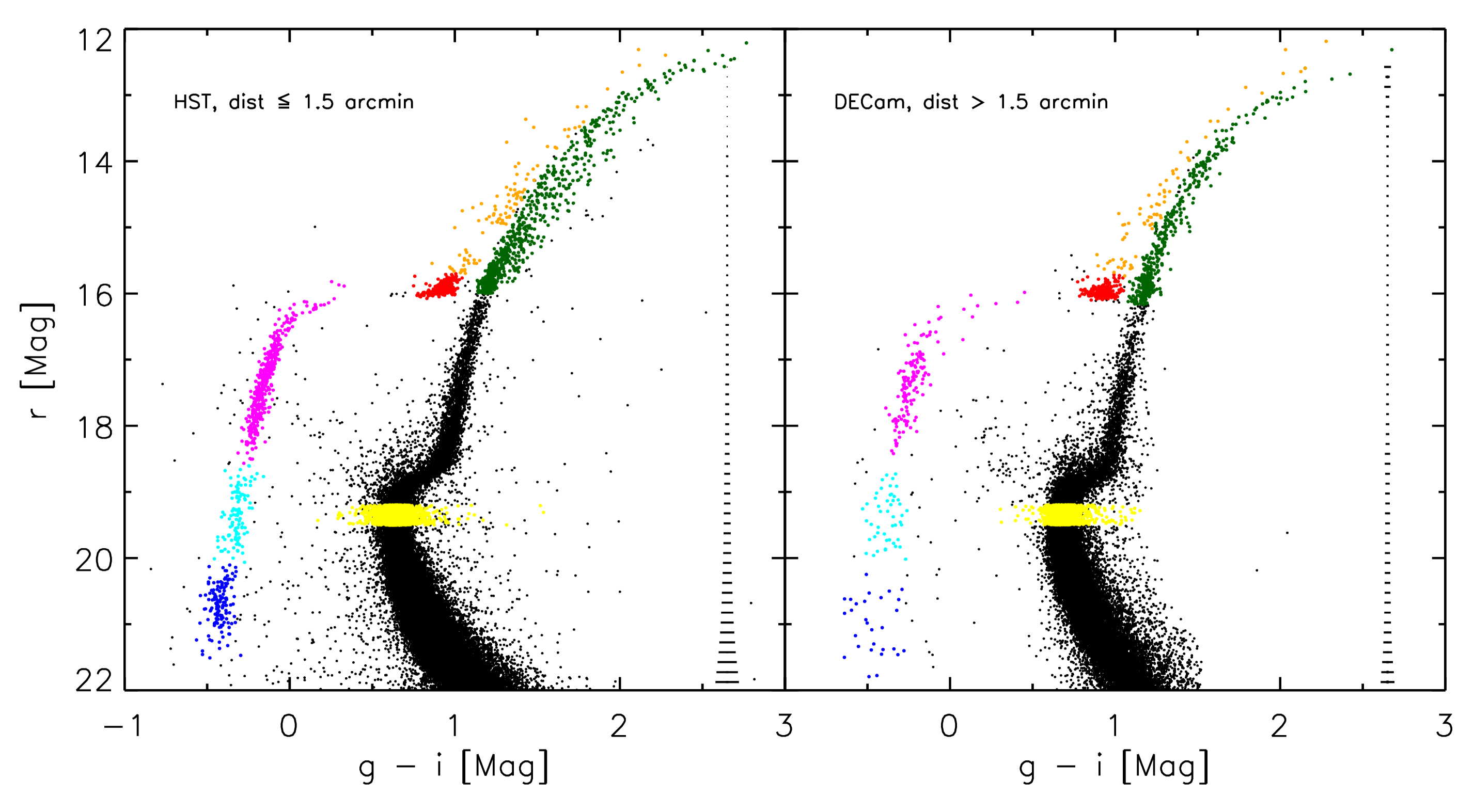} 
\caption{Left -- $r,\ g-i$ CMD based on \emph{HST} photometry for radial distances $\le$ 1.5\arcmin from the center of NGC~2808.  Selected RGB, AGB, MS, and RHB stars are shown as green, orange, yellow, and red filled circles while the blue HB stars, divided into the three groups EBT1, EBT2, and EBT3 groups, are overplotted as magenta, cyan, and blue filled circles, respectively.  Right -- Same but for distances $>$ 1.5\arcmin and based on \emph{DECam} photometry.  Note that the \emph{HST} photometry has been transformed onto the \emph{DECam} photometric system (see text for more details).}
\label{fig:cmd_hb}
\end{figure*}

Fig.\ref{fig:cmd_hb} shows the $r, g-i$ CMDs based on \emph{HST} photometry for radial distances $r \le$ 1.5\arcmin~ (left panel) and \emph{DECam} for $r>$ 1.5\arcmin~ (right).
NGC~2808 HB stars were divided in four groups, namely RHB, EBT1, EBT2, and EBT3, following the prescriptions of \citet{Bedin2000}, \citet{castellani2006}, and \citet{iannicola2009}, and stars were counted; errors were calculated as the square root of the number counts.
Table~\ref{tab:table4} lists the number counts of the HB stars identified in each group and data set with their uncertainties.

\begin{figure}
\includegraphics[width=\columnwidth]{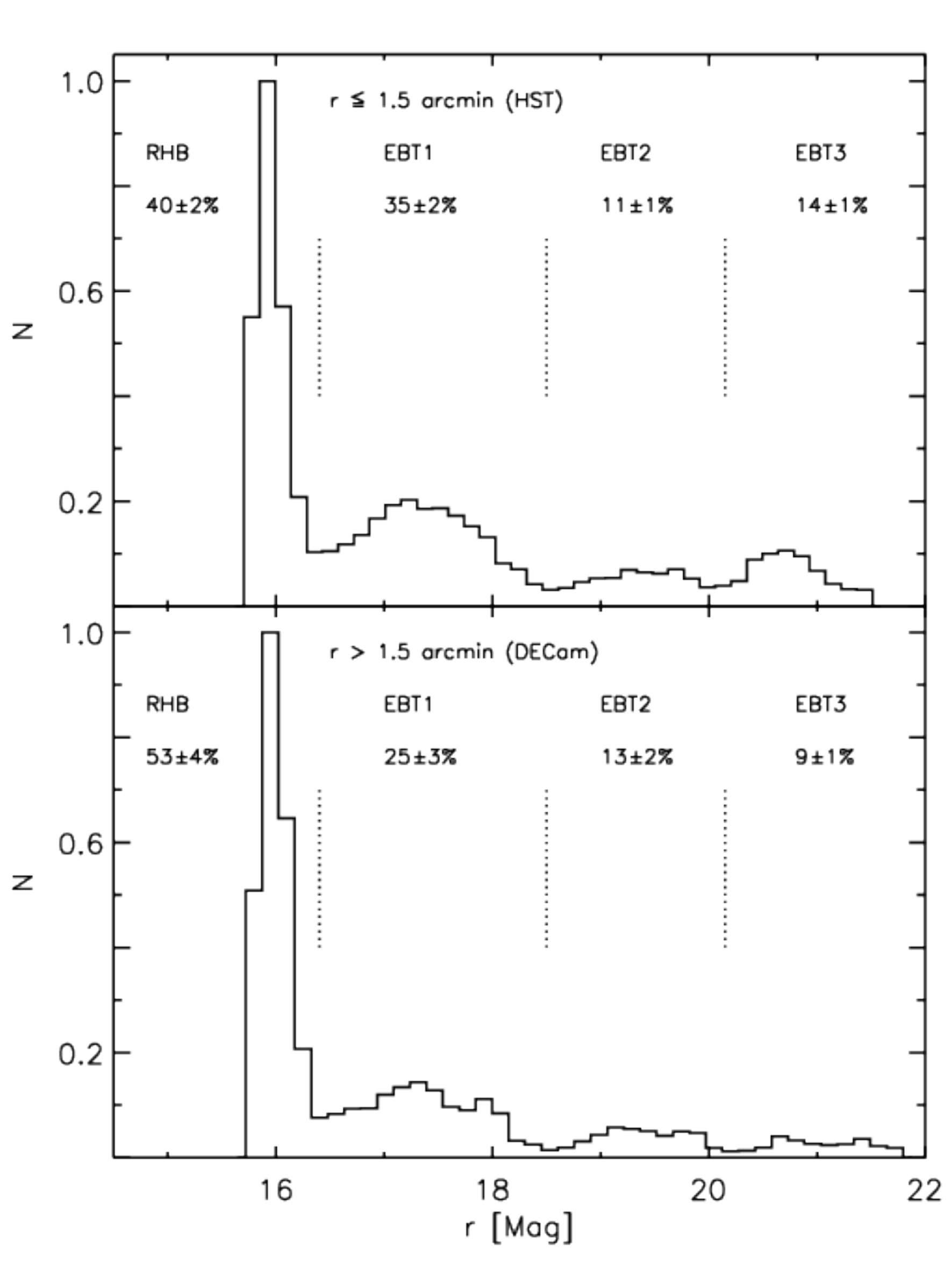}
\caption{Top -- $r$-band luminosity distribution of the different HB groups in NGC~2808, RHB, EBT1, EBT2, and EBT3, respectively, for radial distances from the cluster center $\le$ 1.5\arcmin and based on \emph{HST} photometry. Bottom -- Same but for distances $>$ 1.5\arcmin and based on \emph{DECam} photometry.}
\label{fig:cluster_histo}
\end{figure}

Fig.~\ref{fig:cluster_histo} shows the $r$-band luminosity function for NGC~2808 HB based on \emph{HST} (top panel) and \emph{DECam} photometry (bottom). Note that the completeness of both catalogs is quite similar at these luminosity levels.

It is interesting to note that the RHB fraction increases in the outskirt of NGC~2808, i.e. for distances larger than $\approx$ 1.5\arcmin, i.e. $\approx$ 2$r_h$; on the other hand, the number of EBT1 and EBT3 stars decreases while the number of EBT2 stars is approximately constant. 
However, the \emph{DECam} sample of RHB stars might be more affected by residual contamination of field stars compared to the bluer HB stars. In order to investigate this issue, we matched \emph{DECam} HBs with Gaia DR3 catalog and found 460 (out of 490) stars in common, and 441 of these have a proper motion measurement. Note that only 6 out of 29 EBT3 stars were found in Gaia, due to their faintness. On the other hand, all except two stars of the RHB group were found; therefore, we used the proper motion of NGC~2808 calculated by the \citet[see Table~\ref{table:1}]{Gaia18_GCs}
and selected as candidate cluster members RHB stars with -4 $< \mu_{\alpha} <$ 5 mas/yr and -4 $< \mu_{\delta} <$ 5 mas/yr. Of 266 RHBs, 261 are cluster members, resulting in a $\approx$ 2\% residual contamination of field stars. In the case of the EBT1 star group, all of the 136 stars with Gaia proper motion measurement are cluster members, with a $\approx$ 0\% residual contamination from the field.
We could not perform the same calculation for the EBT2 and EBT3 groups, since these are highly incomplete in the Gaia catalog. However, a large residual contamination by field stars is not expected at these very blue colors, $g -i \lesssim$ 0 mag (see the CMDs in Fig.~\ref{fig:2808_cut}).

We then selected RGB stars brighter than the RHB luminosity level in both $r, g-i$ CMDs, where the RHB is approximately flat for colors $g-i \gtrsim$ 0.5 mag, and calculated the $R$ parameter, i.e. the ratio of the number of HB stars over the number of RGB stars brighter than the RHB, as a function of distance from the cluster center, $r$. Stars were selected in concentric annuli of different thickness, to allow the number of objects per annulus to be always larger than $\approx$ 20. We assumed a Poisson error on the star number counts and we calculated the uncertainty on the ratio as: 

\begin{equation}
\begin{split}
    Err (R) = R \cdot \sqrt{(\frac{dX}{X})^2 + (\frac{dY}{Y})^2} = \\
    R \cdot \sqrt{\frac{1}{X} + \frac{1}{Y}} =  R \cdot \sqrt{\frac{X + Y}{X\cdot Y}}
\end{split}
\end{equation}

where $R$ = $X/Y$ and $X$ and $Y$ are the number of HB and RGB stars in this case,  and $Err (X) = \sqrt{X}$ and the same for $Y$.

\begin{figure}
\includegraphics[width=0.5\textwidth]{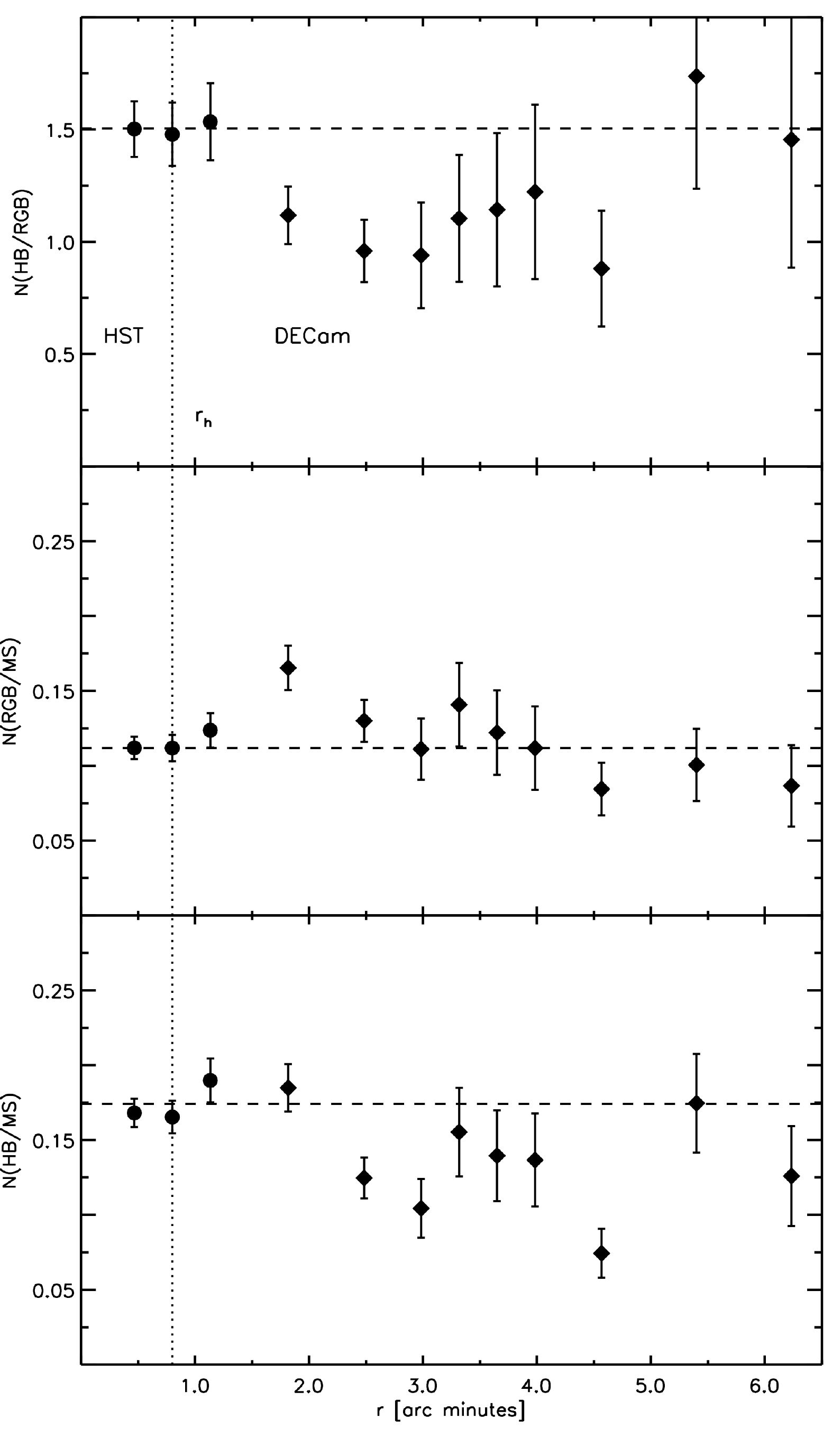}
\caption{Top -- $R$ parameter plotted as a function of distance from the cluster center and based on \emph{HST} and \emph{DECam} star counts. Middle -- Same plot but for the ratio of the number of RGB and MS stars. -- Bottom: Same plot but for the ratio of the number of HB and MS stars.}
\label{fig:rparam}
\end{figure}

The top panel of Fig.~\ref{fig:rparam} shows the $R$ parameter, $N(HB/RGB)$, as a function of radial distance $r$ in arcminutes, based on \emph{HST} photometry for $r \le$ 1.5\arcmin~ and on \emph{DECam} photometry for the external regions. In the case of the \emph{HST} data set we only selected stars for $r \ge$ 0.3\arcmin, since the catalog is not complete closer to the cluster center due to crowding effects. Regarding \emph{DECam} photometry, we calculated the $R$ parameter up to a radial distance of 6\arcmin, since the number of stars greatly decreases for larger distances, with less than 40 RGB and 40 HB stars in the last two annuli, and the uncertainties on the ratios increase.

The $R$ parameter is $\approx$ 1.5 (dashed line in the figure) inside the half-mass radius, $r_h$ (dotted line), while it is systematically lower in the external regions, with a mean value of 1.16$\pm$0.27, or 1.05$\pm$0.25, excluding the last two annuli. For radial distances larger than $\approx$ 5\arcmin, $R$ slightly increases again, but the uncertainty is factor of two larger in these more external regions of NGC~2808.

The decrease of the $R$ parameter with radial distance could be due to an increased contamination of the RGB sample by field stars towards the external regions of NGC~2808. To investigate this issue, we matched the RGB stars with the Gaia DR3 catalog and found 434 (out of 444) stars in common and with a proper motion measurement; of these, 419 are cluster members ($\approx$ 97\%). About half of the contaminant stars (7/15) are located at radial distances larger than 9\arcmin, so outside the range of distances of our analysis (Fig.~\ref{fig:rparam}). Therefore, we can safely claim that contamination by field stars of the RGB sample does not affect the decrease of the $R$ parameter with radial distance.

In order to further assess the cause for the decrease of the $R$ parameter, we also calculated the ratio of the number of HB and RGB stars over the number of MS stars in the same cluster region. We only selected MS stars in a narrow magnitude range around NGC~2808 MS turn-off (MSTO) point, 19.2 $\le r \le$ 19.5 mag, to avoid the number counts being dependent on the mass function of the cluster (see Fig.~\ref{fig:cmd_hb}).

The ratio of the number of RGB and HB stars over the number of MS stars as a function of radial distance is shown in the middle and bottom panels of Fig.~\ref{fig:rparam}, respectively: N(RGB/MS) seems to increase at distances $r \approx$ 1.7\arcmin, and then to decrease attaining a constant value of $\approx$ 0.12 (dashed line in Fig.~\ref{fig:rparam}) throughout the extent of the cluster. On the other hand, the N(HB/MS) is about constant, $\approx$ 0.17 (dashed line), until a radial distance of $\approx$ 2.0\arcmin, and then it slightly decreases to $\approx$ 0.12 at larger distances. 

These ratios suggest that the culprit for the decrease of the $R$ parameter at radial distances larger than 1.5\arcmin~ are the HB stars, in the sense that their total number decreases in the outskirt of NGC~2808.

In order to better understand the trend of the radial distributions of the stars belonging to different evolutionary phases, we also calculated the ratio of the number of AGB stars over the number of RGB stars (brighter than the RHB): N(AGB)/N(RHB) = 0.13$\pm$0.01 for radial distances $<$ 1.5\arcmin, and 0.16$\pm$0.02 at larger distances. The number of AGB stars is also listed in Table~\ref{table:4}. To check for possible field star contamination of the AGB sample in the external regions of NGC~2808, we matched \emph{DECam}   AGB candidates with Gaia DR3 and found 66 stars in common and with proper motion measurements. We applied the same selection criteria as before and found that 88\% of these stars are cluster members. By removing the number of possible field contaminants from both the AGB and RGB samples, the ratio of the number of AGB over the number of RGB stars in the external regions of the cluster is 0.15$\pm$0.02, still slightly larger than the value closer to the cluster center. These ratios would suggest a slight increase of the number of AGB stars in the outer regions of NGC~2808. We then calculated the ratio of the number of AGB stars compared to the number of RHB stars in both regions, and obtained N(AGB)/N(RHB) = 0.21$\pm$0.02 and 0.26$\pm$0.03, respectively. When taking into account the possible contamination by field stars of both samples, the ratio of AGB over RHB stars is 0.23$\pm$0.02 in the external regions of the cluster. These data suggest a similar distribution of AGB and RHB stars, which is expected since most RHBs should evolve along the AGB branch.

We were not able, unfortunately, to compare the star counts of the AGB-manqu\'e stars with those of their progenitors, the blue HB stars, since we do not have precise and deep UV photometry for a wide field of view across NGC~2808 that would allows us to identify the AGB-manqu\'e stars (the bluest \emph{DECam}   filter is the $u$ band with a central wavelength of $\approx$ 3560 \AA). 

\begin{deluxetable*}{lrrrrrrrrr}
\tablecaption{The number of HB stars in the four HB groups, on the RGB and on the MS selected area from the \emph{HST} and \emph{DECam} $r,\ g-i$ CMDs (Fig.~\ref{fig:cmd_hb}) detected in the two different spatial regions of NGC~2808. Figures within brackets give for each region the relative fraction of the different groups with respect to the total number of HB stars. The $R$ parameter values are also listed. \label{table:4} }
\tablewidth{0pt}
\tablehead{
\colhead{Radius}   &
\colhead{N(RHB)}   &
\colhead{N(EBT1)}  &
\colhead{N(EBT2)}  &
\colhead{N(EBT3)}  &
\colhead{N(HB)} &
\colhead{N(RGB)} &
\colhead{N(AGB)} &
\colhead{N(MS)} &
\colhead{$R$}
}
\startdata
$ r \le 1.5$\tablenotemark{a} & 484 (\it 40$\pm$2\%) & 416 (\it 35$\pm$2\%) & 112 (\it 11$\pm$1\%)  & 132 (\it 14$\pm$1\%)  & 1144$\pm$34 & 761$\pm$26 & 102$\pm$10 & 6725$\pm$82 & 1.50$\pm$0.03 \\
$ r > 1.5$\tablenotemark{b}  & 274 (\it 53$\pm$4\%) & 139 (\it 25$\pm$3\%) & 48 (\it 13$\pm$2\%) & 29 (\it 9$\pm$1\%) & 490$\pm$22 & 437$\pm$21 & 71$\pm$8 & 3405$\pm$58 & 1.16$\pm$0.27 \\
Total  & 758$\pm$27 & 555$\pm$23 & 160$\pm$13 & 161$\pm$13 & 1634$\pm$40  & 1198$\pm$35 & 173$\pm$13 & 10130$\pm$101 & \ldots \\
\enddata
\label{tab:table4}
\tablenotetext{a}{Star counts based on \emph{HST} data.}
\tablenotetext{b}{Star counts based on \emph{DECam}   data.}
\end{deluxetable*}

Fig.~\ref{fig:cluster_histo} shows the $r$-band luminosity function of the full HB of NGC~2808 as a function of distance from the cluster center; from this plot it is clear that the number of RHB stars increases in the more external regions of NGC~2808 while the number of EBT1 and EBT3 stars decreases and the EBT2 stars show a flat distribution across the cluster, in agreement with previous results from \citet{Bedin2000} and \citet{iannicola2009}. 


A hypothesis to explain the decrease of EBT3 stars at larger cluster radii is that some fraction of them might originate through the "hot-flasher" scenario; in this case, the stars would be more centrally concentrated as a result of forming in a binary system or as a result of binary interactions \citep{moehler2004, castellani2006}.  \citet{monibidin2011} used GIRAFFE/VLT (ESO) spectroscopy for a sample of hot HB stars (17,000 $\lesssim T_{eff} \lesssim$ 22,000 K) spanning the EBT1 and EBT2 groups to monitor their radial velocities as a sign of binarity.  They found no binaries among the EBT1 group, but found the most probable fractions of close (p < 10 days) and intermediate (p < 10-50 days) period binaries among the EBT2 group to be 20$\%$ and 30$\%$, respectively, thus supporting the "hot-flasher" scenario.  However, we note that this study did not have any EBT3 spectra so its connection to the warmer HB population remains to be confirmed.

Another hypothesis to explain the decrease of EBT1 and/or EBT3 stars toward the outskirts of the cluster is that a fraction of them might be the progeny of a helium-enhanced sub-population in NGC~2808, i.e. the middle or bluest MS; these stars would then be more centrally concentrated as their MS progenitors \citep{dantona2004,simioni2016}



However, a clear correspondence between the different MS and RGB sub-populations and the different HB groups has not been established yet.
\citet{Gratton11} used GIRAFFE/VLT spectroscopy for a sample of 49 RHBs to show that these stars have a similar $O-Na$ anti-correlation as the RGB stars, while the EBT1 stars are mostly $Na$ rich. 

\begin{figure}
\includegraphics[width=0.5\textwidth]{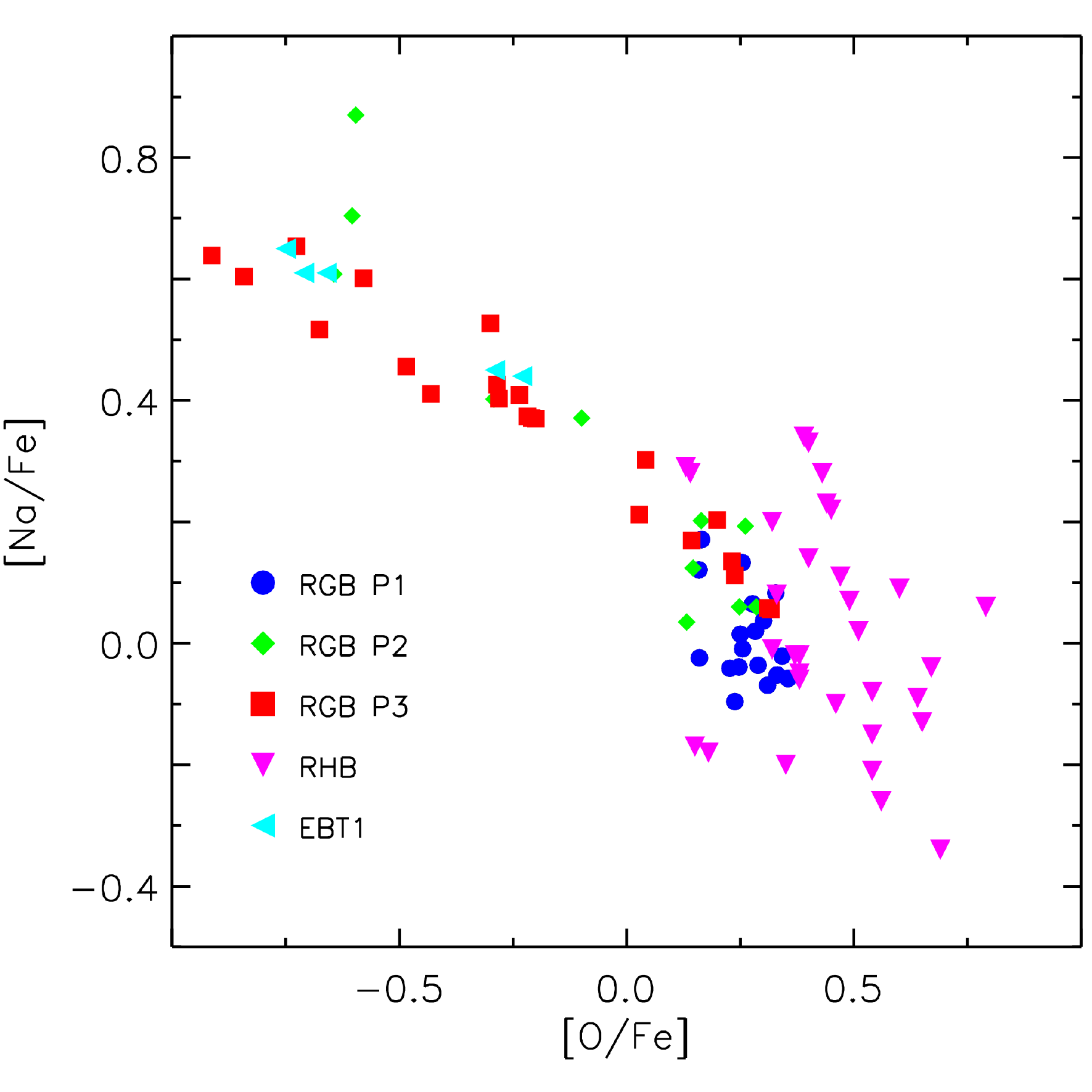}
\caption{$[Na/Fe]$ vs $[O/Fe]$ abundance for RGB stars in common with the spectroscopic study of \citet{Carretta15} divided in our identified three sub-population groups, $P1$ ($Ns$ = 21, blue solid circles), $P2$ ($Ns$ = 13, green diamonds), and $P3$ ($Ns$ = 25, red squares). HB stars in common with the study of \citet{Gratton11} are also over-plotted and divided in RHBs ($Ns$ = 32, magenta down triangles) and EBT1 stars ($Ns$ = 5, cyan horizontal triangles).}
\label{fig:o_na}
\end{figure}

We matched the Gratton et al. spectroscopic data with our photometric catalog and found 37 HB stars in common and with both $Na$ and $O$ abundances measured.  Fig.~\ref{fig:o_na} shows the RGB stars divided into the $P1$, $P2$ and $P3$ groups according to our selection and with $Na$ and $O$ measurement from Carretta in the $[Na/Fe]$ vs $[O/Fe]$ plane. The HB stars with $Na$ and $O$ measurements from Gratton et al. are over-plotted, and the figure shows that RHBs have a spread in the $Na$ and $O$ abundances and cover the same region spanned by the $P1$ and $P2$ RGB sub-populations and some P3 stars.  On the other hand, the EBT1 stars are mostly $Na$ rich and overlap with the $P3$ enriched RGB stars and a few P2 stars on this plane. These data suggest that the RHB stars 
include both the primordial and the light-element enhanced cluster sub-populations, 
while the EBT1 seems to be the progeny of only the more enriched RGB stars. 

It is worth noting that only 5 EBT1 stars have spectroscopic abundances from Gratton et al. and so it is not possible to draw a firm conclusion on the EBT1 group to RGB correspondence. Also, HB stars hotter than $\approx$ 11,500 K (the so-called Grundahl $u$ jump) have atmospheric abundances affected by radiative levitation and diffusion and it is not possible to observe their original $Na$ and $O$ abundances.

From these data, it seems that RHBs are the progeny of more than one stellar sub-populations in the cluster, and possibly all three. Their radial distributions based on the \emph{DECam} + \emph{HST} catalog shows that they are more numerous in the external regions of NGC~2808, with their number increasing by more than 30\% for $r >$ 1.5\arcmin. The AGB stars, which should be the progeny of the RHBs, are also more prevalent in the more external regions. The RHB star spatial distribution is very similar to that of the $P3$ RGB sub-population, whose stars have a more extended spatial distribution for distances larger than 1.5\arcmin. On the other hand, the blue HB stars show an opposite trend, possibly supporting a binary origin.

A similar result for the NGC~2808 RHB was found by \citet{jain2019} using UV photometry collected with UVIT on Astrosat to study the properties of the cluster HB. Their analysis shows that the RHB is composed of at least two different sub-populations and that it has a more extended spatial distribution compared to the bluer HB stars.

   

\section{Discussion and Conclusions}\label{sec:discussion}
We have demonstrated that the \emph{DECam} $C_{ugi}$ pseudo-color index, and its \emph{HST} equivalent, is strongly correlated with an RGB star's light element composition, in particular the $Na$ abundance. These data showed that NGC~2808's RGB population can be decomposed into at least 3 groups, and that these P1, P2, and P3 sub-populations exhibit different spatial density profiles.  The P3 group, the most enriched in light-elements, i.e. $Na$-rich, is the most centrally concentrated inside about one half-mass radius, but becomes much more dispersed in the outer parts of the cluster.  The P3 sub-population centroid is also offset by $\sim$ 6 - 6.5$\arcsec$ from that of the P1 and P2 groups, perhaps indicating a different origin or dynamical evolution.  The P1 and P2 stars exhibit similar spatial profiles, but the more $Na$-enhanced P2 stars are more centrally concentrated out to about 4-5 half-mass radii.  

The stronger central concentration of more $Na$-enhanced RGB stars found in this study is consistent with the previous analysis by \citet{Carretta15}, which found that the intermediate (I) and enhanced (E) RGB stars were more centrally concentrated compared to the primordial (P) stars; however, their spectroscopic sample only covered distances up to 5\arcmin~ from the cluster center. Thanks to our wide-field \emph{DECam} photometric catalog, we were able, for the first time, to show that the more enhanced RGB stars, our P3 group, have a more extended spatial distribution in the outskirt of NGC~2808, almost up to its tidal radius.


Although the $C_{ugi}$ method does not efficiently separate HB stars by their light element compositions, the HB morphology, particularly the blue extent of the HB, is thought to be closely connected to the light element spread of stars in a GGC \citep[e.g.,][]{Carretta07,Gratton10}.  Therefore, we used our deep and precise \emph{DECam} + \emph{HST} photometric catalog to study the spatial distribution of NGC~2808's multi-modal HB. We showed that the relative fraction of RHB stars increases at radial distances $\gtrsim$ 1.5\arcmin~ while the blue HB stars decrease towards the outskirt of the cluster. Moreover, the $R$ parameter, calculated as the ratio of the number of HB stars and the number of RGB stars brighter than the RHB, decreases from a value of $\approx$ 1.5 down to $\approx$ 1.0 at radial distances $\gtrsim$ 2\arcmin.  The ratios of the numbers of HB and RGB stars over a selected sample of MS stars showed that the culprit for the correlated decrease of the $R$ parameter with increasing radial distance is a deficit of HB stars in the outer parts of the cluster.  


The different RGB and HB radial density trends suggest that there is not a direct correspondence between an RGB star's chemical composition and its post-RGB evolution.  For example, the most $Na$-rich RGB stars (P3) constitute 39$\%$ of our sample at r $>$ 1.5$\arcmin$ compared to only 23$\%$ inside 1.5$\arcmin$ from the cluster core.  In contrast, the EBT1 and EBT3 stars have their highest concentrations inside 1.5$\arcmin$ of the core while the EBT2 population either stays the same or increases slightly outside 1.5$\arcmin$.  On the other hand, the RHB fractional contribution increases from 40$\%$ to 53$\%$ when moving from the \emph{HST} ($<$ 1.5$\arcmin$) to \emph{DECam}   ($>$1.5$\arcmin$) sample.  These results are contradictory to the conventional idea of how RGB and HB stars are connected by light element composition (i.e., that more $Na$-rich stars evolve to become bluer HB stars).  However, we can reconcile these patterns if more than one evolutionary path exists for creating the warmest blue HB stars.


As noted in Section \ref{sec:hbrad}, \citet{monibidin2011} found that at least 20-30$\%$ of the EBT2 stars (but 0$\%$ of the EBT1 stars) in NGC~2808 are in close (p $<$ 10 days) or intermediate (p < 10-50 days) binary systems, and their work notes specifically that the warmer blue HB stars could result from both the binary (enhanced mass loss leading to a small HB envelope mass; the "hot flasher" scenario) and the He-enrichment (high original helium abundances pushing the low mass stars to higher HB temperatures) channels.  This scenario implies that the dual formation paths for blue HB stars at least partially erases the expected correlation between $Na$ and $He$ abundance and HB location.  

Fig.~\ref{fig:o_na} shows that while the EBT1 stars correlate with the P3 RGB abundances, the RHB stars exhibit an $O-Na$ anti-correlation that also reaches [Na/Fe] values as high as +0.4 dex, equivalent to the $I2_{C15}$ and some $E_{C15}$ stars from \citet[][see also Fig.~\ref{fig:carretta_comp}]{Carretta15} and the P3 population here.  In other words, the $Na$-enriched P2 and P3 RGB stars can feasibly evolve to occupy almost any HB location.  Furthermore, a recent analysis from \citet{Carlos2022} found at least one very O-poor AGB star in NGC~2808 that could not have evolved from the blue tail of the HB as those stars have masses that are too low to ascend the AGB \citep[e.g., see the discussion in][]{Gratton2010AGB}. 

Further empirical evidence supporting at least some decoupling between RGB composition and HB evolution can be found when examining "second parameter clusters", such as M3 and M13 \citep[see also][]{Lee94}.  In this case, both clusters have similar metallicities, though possibly different ages (e.g., see discussion in \citealt{Gratton10}), along with substantial populations of primordial and enriched stars \citep[e.g.,][see also \citealt{DAntona08}]{Sneden04,JohnsonPilachowski12}, but M13 contains no RHB stars and a long blue tail while M3 exhibits a strong RHB and a much less extended blue HB.  Similarly, NGC~2808 and NGC~6402 are nearly identical in metallicity, age, and mass, both clusters contain stars spanning a wide range of light element abundances, but NGC~2808 has a very prominent RHB and long blue HB while NGC~6402 has almost no RHB stars and a slightly truncated blue HB \citep{Johnson19_6402,DAntona22}.  These two examples highlight that RGB $Na$ and similar abundances alone cannot predict where a star will evolve to on the HB.

From the information above, we posit that our spatial density results for NGC~2808 may be explained if the RHB stars evolved from a mixture of all three RGB populations, the EBT1 stars evolved mostly from more $Na$ and $He$-enhanced (P2 and P3) single stars, and the EBT2 $+$ EBT3 stars formed from a mixture of "hot-flasher" binaries and the most enriched P3 stars.  Following \citet{Dercole08} and \citet{monibidin2011}, the higher masses of the binary systems and initial central formation of $Na$ and $He$-enhanced stars constrains a significant fraction of these groups and their progeny near the cluster core.  At larger radii, mass segregation would cause a decrease in the binary fraction that drives a decline in the EBT2 and EBT3 populations.  However, if the ratio of P3 stars evolving onto the EBT2 versus EBT3 groups is high, then the decline in the EBT2 population fraction with increasing radius would be somewhat mitigated.  Similarly, dynamical interactions might be expected to drive diffusion of the more $Na$ and $He$-enhanced P2 and especially P3 stars, which could be as much as 25$\%$ (0.2M$_{\odot}$) less massive than their "He-normal" counterparts, into the outer parts of the cluster on preferentially radial orbits \citep{Mastrobuono13,bellini20152015,Henault15,Mastrobuono16}.  This could explain the more dispersed density distribution of P3 stars outside the cluster core, and if a significant fraction of such stars can evolve on to the RHB as well as the blue HB then the diffusion process might explain the increasing RHB fraction with increasing radial distance as well.






These results suggest similarities between NGC~2808 and the most massive, and peculiar, GGC \omcen. \citet{Calamida17} and \citet{calamida2020} used deep and precise wide-field \emph{DECam} photometry combined with \emph{HST} data for the cluster center to show that the most metal-rich RGB stars are centrally concentrated but have a more extended spatial distribution compared to the more metal-poor RGB stars; also, their centroid is shifted by $\approx$ 1\arcmin (0.28r$_{h}$) relative to the centroid of the primordial metal-poor stellar population. Blue MS stars are also more centrally concentrated with a more extended spatial distribution compared to the red MS.

Regarding the HB, \citet{castcal2007} showed that the bluest HB stars in \omcen (EBT3, or EHBs, note that \omcen does not have a RHB) are more centrally concentrated, supporting their origin as hot flashers. This was further confirmed by the spectroscopic analyses of \citet{moehler2007,moehler2011} and \citet{latour2014}, which showed EHB stars cannot only be the progeny of the supposedly helium-enhanced sub-population in \omcen since a large fraction of them also show $C$-enhancement in their atmosphere, resulting from the mixing between the helium- and carbon-rich core and the hydrogen envelope \citep{sweigart1997, brown2001, miller2008, cassisi2009}. On the other hand, a radial velocity study of 152 EHB stars in \omcen showed a close-binary fraction of only $\approx$ 5\% \citep{latour2018}. However, this analysis was not sensitive to intermediate- and long-period binaries, which could be the majority of the EHB binaries, such as \citet{monibidin2011} has shown in NGC~2808.

The analyses presented here and in \citet{Calamida17,calamida2020} highlight the power of extending population investigations beyond a few half-light radii with wide-field imagers, especially for massive GGCs.  Radial density trends observed in small \emph{HST} fields are not necessarily representative of a cluster's global properties, and we do not yet have a clear enough understanding about post-RGB evolution to fully link the different groups identified in MS, RGB, HB, and AGB analyses.  However, renewed investigations into the outer parts of clusters, where dynamical evolution times are longer, with space-based UV imaging, ground-based wide-field imaging (e.g., \emph{DECam}   $C_{ugi}$), and spectroscopic abundance analyses and radial velocity monitoring should provide new insight into the complicated formation histories of the Galaxy's globular clusters.

\section*{Acknowledgements}
C.A.P. acknowledges the generosity of the Kirkwood Research Fund at Indiana University.  AMB acknowledges funding from the European Union’s Horizon 2020 research and innovation programme under the Marie Skłodowska-Curie grant agreement No 895174.  This study was supported by the DDRF grant D0001.82481. This project used data obtained with the Dark Energy
Camera (DECam), which was constructed by the Dark Energy Survey (DES)
collaboration.  Funding for the DES Projects has been provided by the U.S.
Department of Energy, the U.S. National Science Foundation,
the Ministry of Science and Education of Spain, the Science and Technology
Facilities Council of the United Kingdom, the Higher Education Funding Council
for England, the National Center for Supercomputing Applications at the
University of Illinois at Urbana-Champaign, the Kavli Institute of Cosmological
Physics at the University of Chicago, the Center for Cosmology and
Astro-Particle Physics at the Ohio State University, the Mitchell Institute for
Fundamental Physics and Astronomy at Texas A\&M University, Financiadora de
Estudos e Projetos, Funda{\c c}{\~a}o Carlos Chagas Filho de Amparo {\`a}
Pesquisa do Estado do Rio de Janeiro, Conselho Nacional de Desenvolvimento
Cient{\'i}fico e Tecnol{\'o}gico and the Minist{\'e}rio da Ci{\^e}ncia,
Tecnologia e Inovac{\~a}o, the Deutsche Forschungsgemeinschaft,
and the Collaborating Institutions in the Dark Energy Survey.  The
Collaborating Institutions are Argonne National Laboratory,
the University of California at Santa Cruz, the University of Cambridge,
Centro de Investigaciones En{\'e}rgeticas, Medioambientales y
Tecnol{\'o}gicas-Madrid, the University of Chicago, University College London,
the DES-Brazil Consortium, the University of Edinburgh,
the Eidgen{\"o}ssische Technische Hoch\-schule (ETH) Z{\"u}rich,
Fermi National Accelerator Laboratory, the University of Illinois at
Urbana-Champaign, the Institut de Ci{\`e}ncies de l'Espai (IEEC/CSIC),
the Institut de F{\'i}sica d'Altes Energies, Lawrence Berkeley National
Laboratory, the Ludwig-Maximilians Universit{\"a}t M{\"u}nchen and the
associated Excellence Cluster Universe, the University of Michigan,
{the} National Optical Astronomy Observatory, the University of Nottingham,
the Ohio State University,
the OzDES Membership Consortium
the University of Pennsylvania,
the University of Portsmouth,
SLAC National Accelerator Laboratory, Stanford University,
the University of Sussex,
and Texas A\&M University.  Based on observations at Cerro Tololo
Inter-American Observatory, National Optical Astronomy Observatory
(2013A-0529;2014A-0480; R.M. Rich), which is operated by the Association of
Universities for Research in Astronomy (AURA) under a cooperative agreement
with the National Science Foundation.  This work has made use of data from the
European Space Agency (ESA) mission {\it Gaia}
(\url{https://www.cosmos.esa.int/gaia}), processed by the {\it Gaia}
Data Processing and Analysis Consortium (DPAC,
\url{https://www.cosmos.esa.int/web/gaia/dpac/consortium}). Funding for the
DPAC has been provided by national institutions, in particular the institutions
participating in the {\it Gaia} Multilateral Agreement.

%

\vspace{5mm}
\facilities{CTIO:4.0m, HST:WFC3, HST:ACS}

\bibliographystyle{aasjournal}
\bibliography{references}

\end{document}